\documentclass[preprint]{aastex}

\usepackage{epsfig}
\usepackage[]{natbib}

\newcommand{\eg}{{\rm e.g.,}}
\newcommand{\ie}{{\rm i.e.,}}

\newcommand{\Mstar}{\ensuremath{M_{\star}}}

\newcommand{\Lstar}{\ensuremath{L_{\star}}}

\newcommand{\Leff}{\ensuremath{L_{\rm eff}}}

\newcommand{\amin}{\ensuremath{\arcmin}}
\newcommand{\asec}{\ensuremath{\arcsec}}

\newcommand{\kmsMpc}{~\ensuremath{{\rm km\ s}^{-1}\ {\rm Mpc}^{-1}}}

\newcommand{\hkpc}{~\ensuremath{{h}^{-1}\ {\rm kpc}}}
\newcommand{\hMpc}{~\ensuremath{{h}^{-1}\ {\rm Mpc}}}

\newcommand{\Hnought}{\ensuremath{{\rm H}_0}}
\newcommand{\znought}{\ensuremath{z_0}}

\newcommand{\omegam}{\ensuremath{\Omega_{\rm m}}}
\newcommand{\omegal}{\ensuremath{\Omega_{\lambda}}}
\newcommand{\omegak}{\ensuremath{\Omega_{k}}}

\newcommand{\wt}{\ensuremath{\omega(\theta)}}
\newcommand{\Aw}{\ensuremath{A_{\omega}}}
\newcommand{\xr}{\ensuremath{\xi(r)}}
\newcommand{\xrz}{\ensuremath{\xi(r,z)}}
\newcommand{\rnought}{\ensuremath{r_0}}
\newcommand{\xnought}{\ensuremath{x_0}}

\def \figwidth {\linewidth}

\shorttitle{Title}
\shortauthors{Wilson \etal}

\begin{document}

\title{GALAXY CLUSTERING EVOLUTION IN THE UH8K WEAK LENSING FIELDS\altaffilmark{1}}

\author{Gillian Wilson\altaffilmark{2,3}}

\email{gillian@ipac.caltech.edu}

\altaffiltext{1}{Based on observations with the Canada-France-Hawaii Telescope which is operated by the National Research Council of
Canada, le Centre National de la Recherche Scientifique de France, and the University of Hawaii.}

\altaffiltext{2}{Physics Department, Brown University, 182 Hope Street, Providence, RI 02912}
\altaffiltext{3}{SIRTF Science Center, California Institute of Technology, 220-6, Pasadena, CA 91125}

\begin{abstract}
We present measurements of the two-point galaxy angular correlation function
as a function of apparent magnitude, color, and morphology. Our
analysis utilizes
images taken using the UH8K CCD mosaic camera on the CFHT.
Six $0\degr5 \times 0\degr5$
fields were observed for a total of 2 hours each in $I$ and $V$,
resulting in catalogs containing $\sim 25 000$ galaxies per field.
We present new galaxy number counts to limiting 
magnitudes of $I=24.0$ and $V=25.0$.
We divide each passband sample into intervals of width one magnitude.
Within each magnitude interval, we 
parameterize the angular correlation function by  $A_w\theta^{-\delta}$,
and find \wt\ to
be well described by a power-law of index $\delta=0.8$. We find the amplitude
of the correlation function, $A_w$, to decrease monotonically with 
increasingly faint apparent magnitude. 
We compare with predictions which utilize redshift distributions based
on deep spectroscopic observations.
We conclude that simple redshift-dependent models which characterize
evolution by means of the $\epsilon$ parameter 
inadequately describe the observations. This is because
the predictions do not allow for the varying mix of morphologies and absolute
luminosities (and hence clustering
strengths) of galaxies sampled at different apparent magnitudes.
We find a strong clustering dependence on $V-I$ color. This is  because galaxies of
extreme color lie at similar redshifts and the angular correlation functions for these samples
are minimally diluted by chance projections. We find 
extremely red $(V-I=3.0)$ galaxies 
(likely early-type galaxies at $z\sim1$) to have an \Aw\ about $10$ times, and
extremely blue ($V-I=0.5$) galaxies (likely local late types)
to have an \Aw\ about $15-20$ times
higher than that measured for the full field population.

We then present the first attempt to investigate the redshift
evolution of clustering,
utilizing a population of galaxies of the \emph{same} morphological
type and absolute luminosity.
We study the dependence of \wt\ on redshift for \Lstar\
early-type galaxies in
the redshift range $0.2<z<0.9$. Although uncertainties are large,
we find the evolution in the clustering of these galaxies to be
consistent with stable clustering [$\epsilon=0$ for a redshift
dependence of the spatial correlation function, \xr, parametererized
as $\xrz = {\left( {r}/{\rnought} \right)}^{-\gamma} (1+z)^{-(3+\epsilon)}$].
We find  \Lstar\ early-type galaxies to cluster slightly 
more strongly (physical correlation length $\rnought = 5.25\pm0.28 \hMpc$
assuming $\epsilon=0$) than the local full 
field population.
This is in good agreement with the correlation
length measured by the 2dFGRS 
for \Lstar\ early-type galaxies in the local universe.

\end{abstract}


\section{INTRODUCTION}
\label{sec:intro}

One of the most interesting and important problems in modern astronomy is 
that of galaxy formation and evolution. Traditionally, galaxy
distributions have been quantified using correlation functions.
Two principle approaches, each
with its own advantages and drawbacks, 
have been used to measure the two-point function, which measures the excess
probability over random 
of finding another galaxy at a separation
$r$ from a given galaxy.  One option is to
compute the spatial correlation function, $\xr$, or its Fourier
transform equivalent $P(k)$, directly using
spectroscopic redshifts.
In recent years a plethora of redshift surveys culminating in the
2-degree Field Galaxy Redshift Survey
(hereafter 2dFGRS) and Sloan Digital Sky Survey (hereafter SDSS) have been
providing accurate pictures of the distribution of galaxies in the local universe.
The spectroscopic approach, however, is limited by the technical difficulties
inherent in measuring spectra for many galaxies and the unfeasibility of
obtaining both a very deep and very wide sample. 

In order to quantify the clustering of faint galaxies beyond spectroscopic
limits the angular two-point correlation function, \wt, has been the approach of choice.
The power of \wt\ as a diagnostic of the galaxy distribution lies in
the simplicity of its application; one is required only to count pairs of
galaxies at given angular separations and normalize the results with
respect to the number of pairs expected from a random distribution.
The drawback is that angular correlation function analyses must rely on 
accurate observations or models of the redshift
distribution of faint galaxies to invert \wt, and hence
deduce the three-dimensional correlation length, \rnought.

Pioneering studies of \wt\ using photographic plates were
carried out by \citet{grothp-77, phil-78, sfem-80, mad-90, bern-94} and \citet{ip-95},
among others. In the last decade or so the
widespread use
of charge-coupled device (CCD) cameras have permitted studies reaching 
far deeper limiting magnitudes
[passbands in parenthesis]  \eg\ 
\citet{efs-91} [$U$,$B$,$R$,$I$]; 
\citet{roche-93} [$B$,$R$];
\citet{bsm-95} [$R$];        
\citet{hudlil-96} [$R$];
\citet{lp-96} [$I$];
\citet{vill-97} [$R$];       
\citet{wf-97} [$V,R,I$];     
\citet{bs-98} [$I$];         
\citet{post-98} [$I$];       
\citet{cabanac-00} [$V$,$I$]; 
\citet{fynbo-00} [$R$,$I$];   
\citet{mcc-00} [$B$,$R$,$I$,$K$]. 
These investigations were able only to target one or two
fields up to about $50$ arcmin$^{2}$ in size.
Several groups attempted to cover larger areas by
mosaicing together many separate pointings 
(\citeauthor{post-98}; \citet{re-99} [$R$]) but
these observations reached much shallower limiting magnitudes.
In recent
years, the advent of wide-field mosaic cameras on 4-m class telescopes
has begun to revolutionize the field, permitting studies of unprecedented
depth and areal coverage, with the corresponding reduction in variance 
inherent in covering large areas of the sky
(\citeauthor{cabanac-00}; \citet{mcc-01} [$I$]).

Spatial two-point correlation functions for local, bright,
optically selected samples have been
determined by many authors over the last few decades.
Numerous studies have found \xr\ to be well described
by a power law, $\xr=(r/ \rnought)^{-\gamma}$, with slope
$\gamma\simeq1.8$
and correlation
length $\rnought\simeq5 \hMpc$ 
for $r \lesssim 15 \hMpc$ 
\citep{dp-83, love-95, norb-01}.
$\rnought^{\gamma}$ may be interpreted as the correlation amplitude at $1 \hMpc$.
Thus the value of \rnought\ provides a measure of the clustering
strength of galaxies in the sample, with a larger value implying stronger clustering.

In the last few years, it has become generally accepted
that clustering strengths, as measured in the local universe, 
have a dependence on galaxy morphology
\citep{iov-93, love-95, hermit-96, willmer-98, norb-02}. It has been known since at least \citet{dg-76}
that early-type galaxies (ellipticals and S0's) cluster more strongly
than late types \ie\ \rnought$({\rm E}) >$ \rnought$({\rm S})$.
Estimates of the ratio of the correlation strengths
$[$\rnought$({\rm E})$/\rnought$({\rm S})]^{1.8}$
for the two types range from $\sim1.2$ to $\sim5$.
Claims for a luminosity dependent component to galaxy correlation strengths
have been more
contentious, although results from the most recent surveys
\citep{hermit-96, lin-96, willmer-98, guzzo-00, norb-01} do indeed
seem to 
indicate that high-luminosity galaxies are more strongly clustered 
than low-luminosity galaxies. 
A dependence of clustering strength on both intrinsic luminosity and
morphological type is to be expected
if galaxies are biased tracers (see below) of the mass distribution in the
universe.
Early efforts to quantify galaxy evolution as a function of redshift 
utilizing only \emph{one} passband  proved rather limiting. This is because  
subsets of any given sample selected on apparent magnitude contain
galaxies of differing absolute luminosities (and differing morphological types)
at different redshifts, greatly complicating the analysis. Whenever two
(or more) passbands
have been available,  color-selection has often been used.
\citet{ip-95, nw-95, lsk-96, rrgim-96, wf-97, brown-00, cabanac-00,
  mcc-01} and  \citet{zehavi-02} 
all found red galaxies to cluster more strongly than blue
galaxies. This is most likely a manifestation of the morphological
clustering dependence.
A more recent innovation has been the use of multi-passband data to
estimate photometric redshifts \citep{conn-98, arnc-99, brun-00, tep-01}, 
a technique which bridges traditional
spectroscopic and photometry camps to study galaxies in the
range $0<z\leq1$. However, although promising, to date, photometric redshifts analyses
have been limited to small fields-of-view. 

The ``$\epsilon$'' formalism, first introduced by \cite{grothp-77}, has
traditionally been used to characterize the evolution of clustering 
with redshift. This empirical approach assumes that the typical
clustering length observed at high redshift transitions monotonically  to that
observed locally. By assuming a redshift distribution and cosmology, 
it is possible to predict \wt, compare to observations and hence
determine the value of $\epsilon$ which best describes the evolution
of the correlation function. 
Many authors have concluded, either from the aforementioned \wt\ studies or
directly from spectroscopic studies 
\citep{lefevre-96, shep-97, small-99, carl-00, hcb-00},
that $0<\epsilon<2$ \ie\ that galaxy clustering is either stable or 
grows in amplitude from $z=1$ to the present.
Undoubtedly, a lack of consistency of sample \ie\
an inability to follow the \emph{same} population of galaxies
due to changing morphological mixes and intrinsic luminosities with redshift,
makes interpretation of the measurements confusing.

Strong evidence that the ``$\epsilon$'' formalism might be invalid, 
at least at higher redshift, was provided by the discovery that
 clustering amplitudes measured for Lyman-break
galaxies at $z\sim3$ are similar to those measured for local galaxies
\citep{adel-98, giav-98}.
Such strong clustering of the Lyman-break galaxies is to be
expected if these galaxies are highly
biased with respect to the mass distribution. 
In the standard hierarchical picture of galaxy formation and evolution,
galaxy clustering traces overdense regions in the dark matter
distribution. High redshift galaxies are expected to form at the
most extreme peaks in the density field and are thus biased tracers of the
mass \citep{kais-84,bbks-86}. If early-type or 
more luminous galaxies are associated with rarer, more massive halos, then 
these galaxies would be expected to exhibit even stronger clustering than 
the galaxy population as a whole.
Subsequent to the epoch of formation, the clustering of 
the galaxies is expected to  
evolve more slowly than the clustering of the dark matter
so the two distributions are expected to be more similar today than in the past
\citep{baugh-99, kauffhiz-99, carl-00}. 

In the hierarchical formation scenario, the clustering of dark matter
increases monotonically with time. The rate of evolution in the
clustering of the dark
matter is a function of cosmology,
being faster in a high density ($\epsilon\sim1.0$ for $\Omega_{m}=1$) than 
in a low density ($\epsilon\sim0.2$ for $\Omega_{m}=0.2$) universe \citep{colin-97}.
If the evolution in 
dark matter clustering could be measured directly as a function of
redshift then $\omegam$, $\omegal$, and the power spectrum of 
density fluctuations could be inferred directly.
However, in practice, one can measure only evolution in the galaxy clustering pattern.
The amplitude of the galaxy correlation function is determined by
a combination of factors: evolution in the underlying dark matter
fluctuations plus any
bias (which is also a function of cosmology) 
relating the galaxy overdensities to the mass.
Realistically, in addition to the depth of the dark matter potential,
one can expect galaxy evolution
also to depend on complex physical processes 
involving the local environment, cooling and feedback mechanisms, and galaxy
interactions. 
Observations of galaxy clustering at high redshift are therefore vital to
constrain models of these processes empirically, and deepen
our understanding of the complexities of both galaxy and structure
formation and evolution.

In this paper we investigate galaxy clustering evolution
on scales of up to $30\amin$ using data collected using the University of
Hawaii's 8K (UH8K) CCD mosaic camera on the Canada-France-Hawaii
Telescope (CFHT).
The data were obtained for a weak lensing
study of ``blank fields'' \ie\ the fields chosen for study were intended to
be representative views of the universe not 
containing any unusually large masses such as rich clusters. 
The data are also, therefore, well suited to the study of (field) galaxy
clustering and evolution.

The outline of the paper is as follows.  In \S\ref{sec:data} we describe the
data and photometry. 
In \S\ref{sec:counts} we present both our number counts and a comparison with the
counts from other groups. In \S\ref{sec:wt} we describe and discuss our measurements
of the two-point correlation function as a function of median apparent
magnitude,
$V-I$ galaxy color, and morphological type. We compare to predictions
and discuss possible sources of uncertainty.
In \S\ref{sec:conc}  we briefly summarize our conclusions.
We assume a 
flat lambda ($\Omega_{{\rm m}0} = 0.3, \Omega_{\lambda 0} = 0.7$) 
cosmology with $\Hnought = 100 $ $h\kmsMpc$ throughout.

\section{THE DATA AND GALAXY SAMPLES}
\label{sec:data}

\subsection{Data Acquisition and Reduction}

The data were taken at the 3.6m CFHT telescope using the $8192 \times 8192$
pixel UH8K camera at prime focus. The field of view of this camera is 
$\sim 30 \amin$ with pixelsize $0.207$\asec. The data (six pointings) used in the analysis were acquired as part
of an ongoing 
project that has the principle aim of investigating the cosmic shear pattern caused
by gravitational lensing from the large-scale structure  of the universe.
In addition to the main project, estimates of cosmic shear variance 
on $2' - 30'$ scales \citep*[Paper I]{kwl-00}, the data have also been
used to 
investigate galaxy halos 
at radii of $20'' - 60''$ ($50 -200\hkpc$) \citep*[Paper II]{wklc-01}, and to investigate the relationship 
between mass and light on group and cluster scales ($45'' - 30'$) \citep*[Paper III]{wkl-01}.
In this paper we focus on properties of galaxy
clustering.

Table~\ref{tab:fields} gives an overview of the data, describing the field 
name, center and seeing for each pointing. The reduction procedure was
lengthy and involved. We defer a full description of the data reduction
pipeline (involving careful registration and point-spread function 
correction) and the resulting catalogs to a later paper
\citep{wkcats-03}. Here we provide only an overview
of the photometry pipeline.

\subsection{Photometry}

The data were dark subtracted, flat-fielded, registered, median averaged
and corrected for galactic extinction using extinction measurements
from \citet*{schl-98}. The UH8K science images were calibrated 
to the Johnson-Cousins system
using a series of standard star field observations \citep{lan-92}.
The standard star fields were also dark subtracted and flat-fielded
in a similar manner to the science frames. These were used then to calculate
zero-points for each pointing. In agreement with \cite{mcc-01} our
observations did not indicate the presence of a color term for either the
$I$ or $V$ filters. 

The imcat data reduction package
\url({http://www.ifa.hawaii.edu/$\sim$kaiser/}) was used throughout. Objects
were detected and assigned an optimal size 
by smoothing the science frames by a progressively larger series
of filters as described in \citet{ksb-95}. Aperture magnitudes were then measured
within a multiple of three times this radius.

The number counts of objects at faint magnitudes are dominated by
galaxies but at brighter magnitudes the stellar component is
dominant. We removed
stars at brighter magnitudes ($I\lesssim23$ ; $V\lesssim24$) by hand, by means of filtering on a size-magnitude diagram. For fainter objects, no  attempt was made to further
eliminate stars from the sample since compact galaxies could be
mistakenly removed and 
stellar numbers are very small relative to the galaxies at these faint limits.
After stellar filtering, approximately $25000$ galaxies remained 
in each passband for
each of the six pointings, an extremely deep and wide-area dataset
compared to previous studies.

\section{NUMBER COUNTS}
\label{sec:counts}

\begin{figure}
\centering\epsfig{file=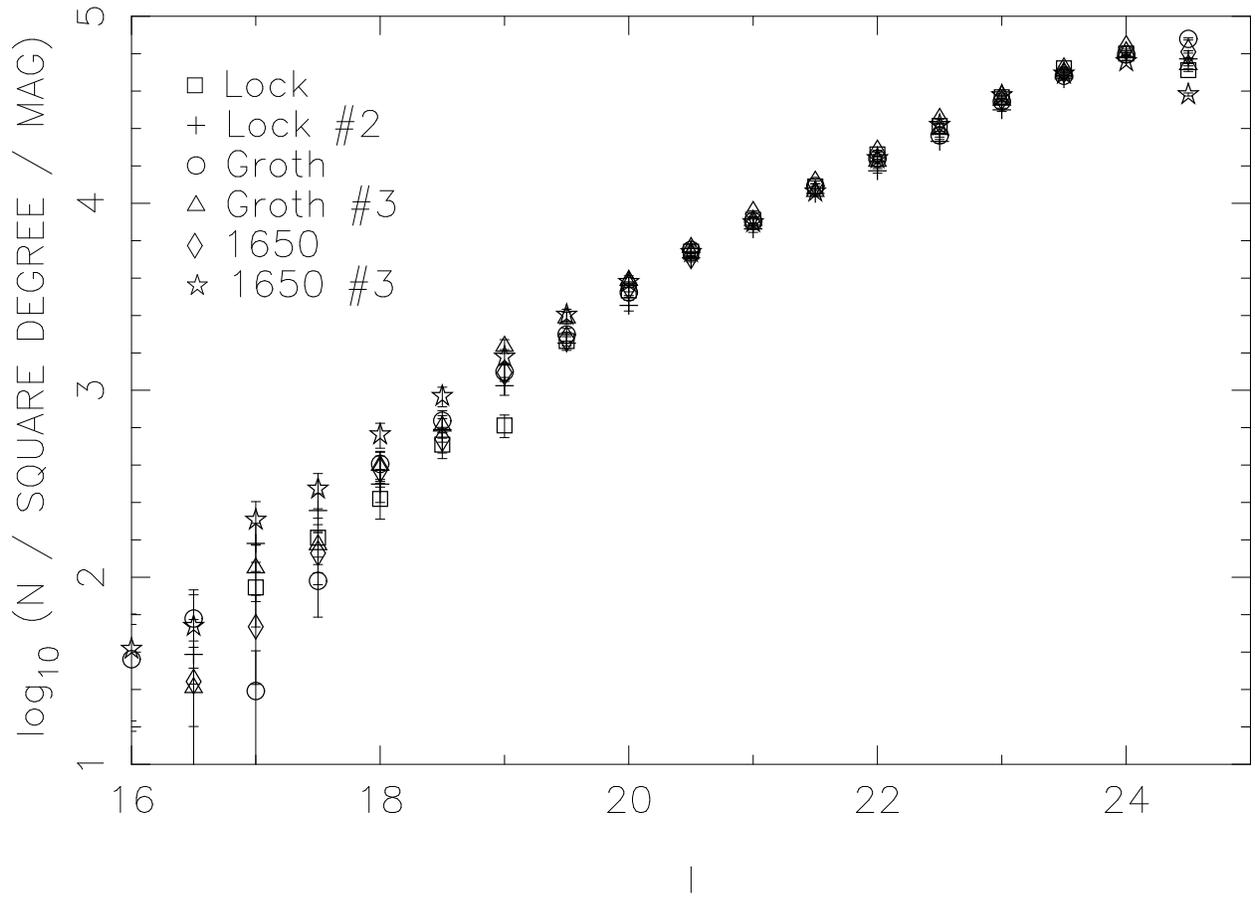,width=\figwidth}
\caption[countsi.ps]{
$I$-band number counts. 
The six pointings are as indicated by the key. 
 \label{fig:countsi}
}
\end{figure}

\begin{figure}
\centering\epsfig{file=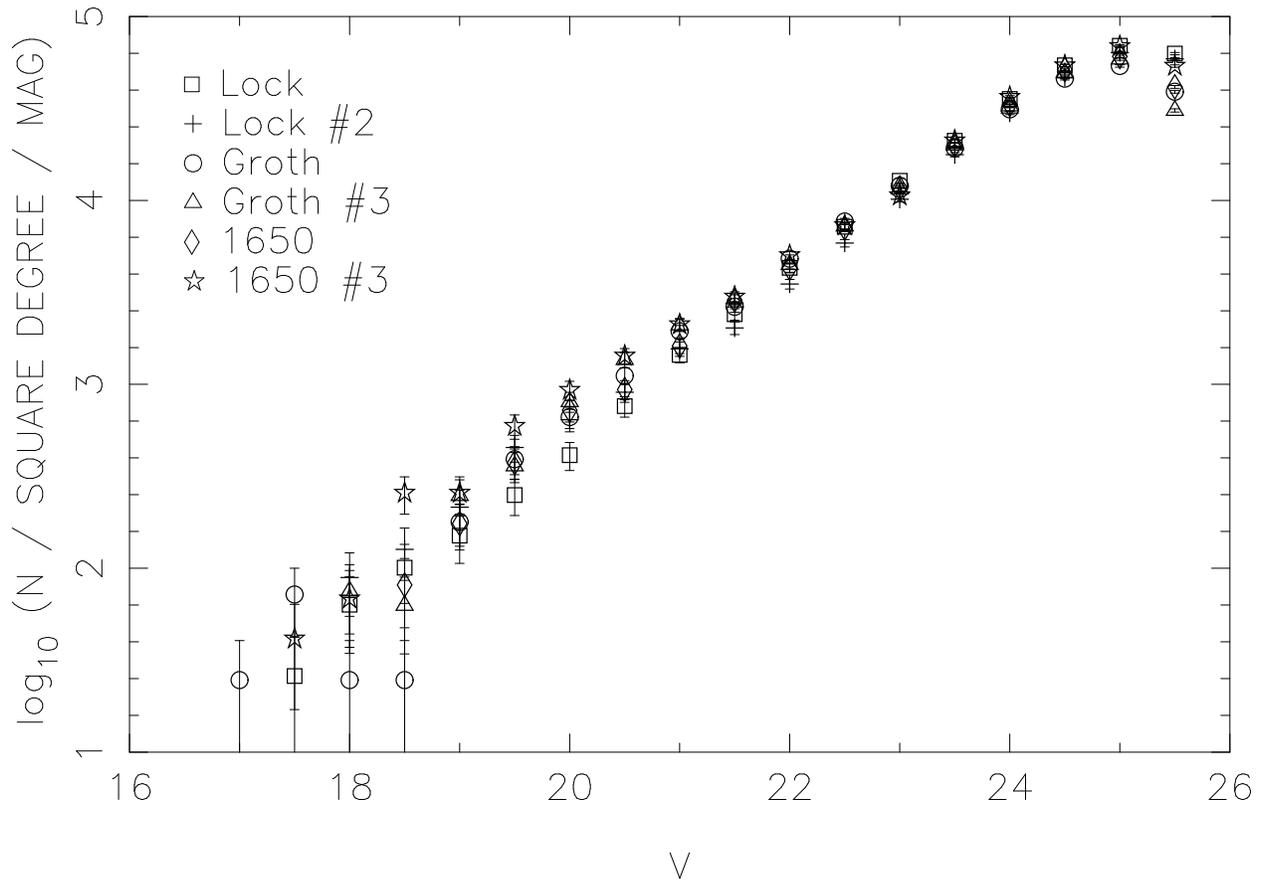,width=\figwidth}
\caption[countsv.ps]{
Same as Fig.~\ref{fig:countsi} but for $V$-band.
\label{fig:countsv}
}
\end{figure}

The $I$-band number counts (logarithm of number of galaxies 
degree$^{-2}$ mag$^{-1}$) for each pointing are shown in Figure~\ref{fig:countsi}.
The uncertainties shown are Poissonian. Clearly, the counts from each pointing
are in good agreement with each other and are complete to $I=24.0$.
Figure~\ref{fig:countsv} shows the same but for the $V$-band counts
which are complete to 
$V=25.0$.

\begin{figure}
\centering\epsfig{file=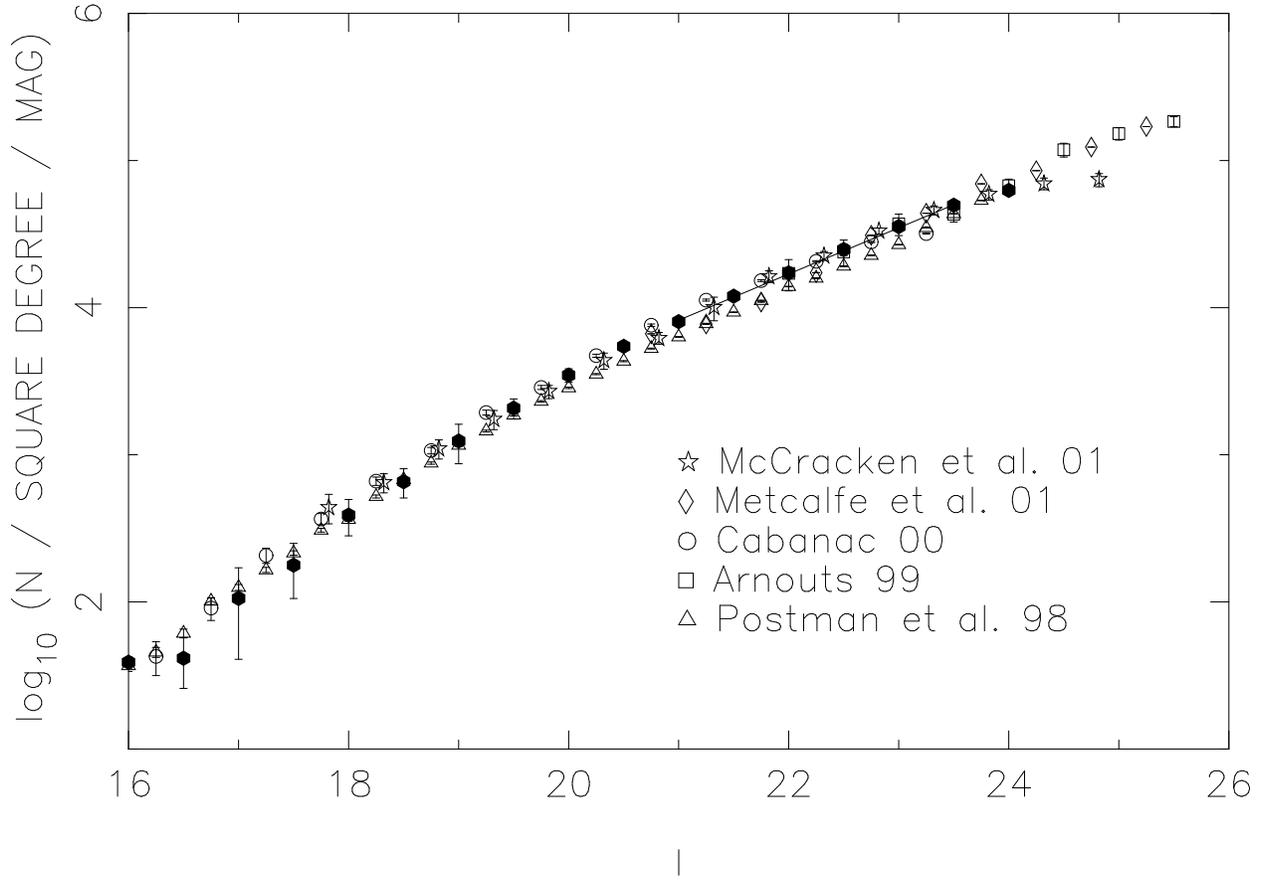,width=\figwidth}
\caption[countsi_all.ps]{
Comparison of our $I$-band number counts with measurements from other groups. 
The filled hexagons represent our data. The uncertainties are obtained from
the pointing-to-pointing variations in Figure~\ref{fig:countsi}. See Table~\ref{tab:countsi} for values.
The line shows the best fit to our counts in the range $21.0<I<23.5$ (slope of $0.31\pm0.01$). 
 \label{fig:countsi_all}
}
\end{figure}

\begin{figure}
\centering\epsfig{file=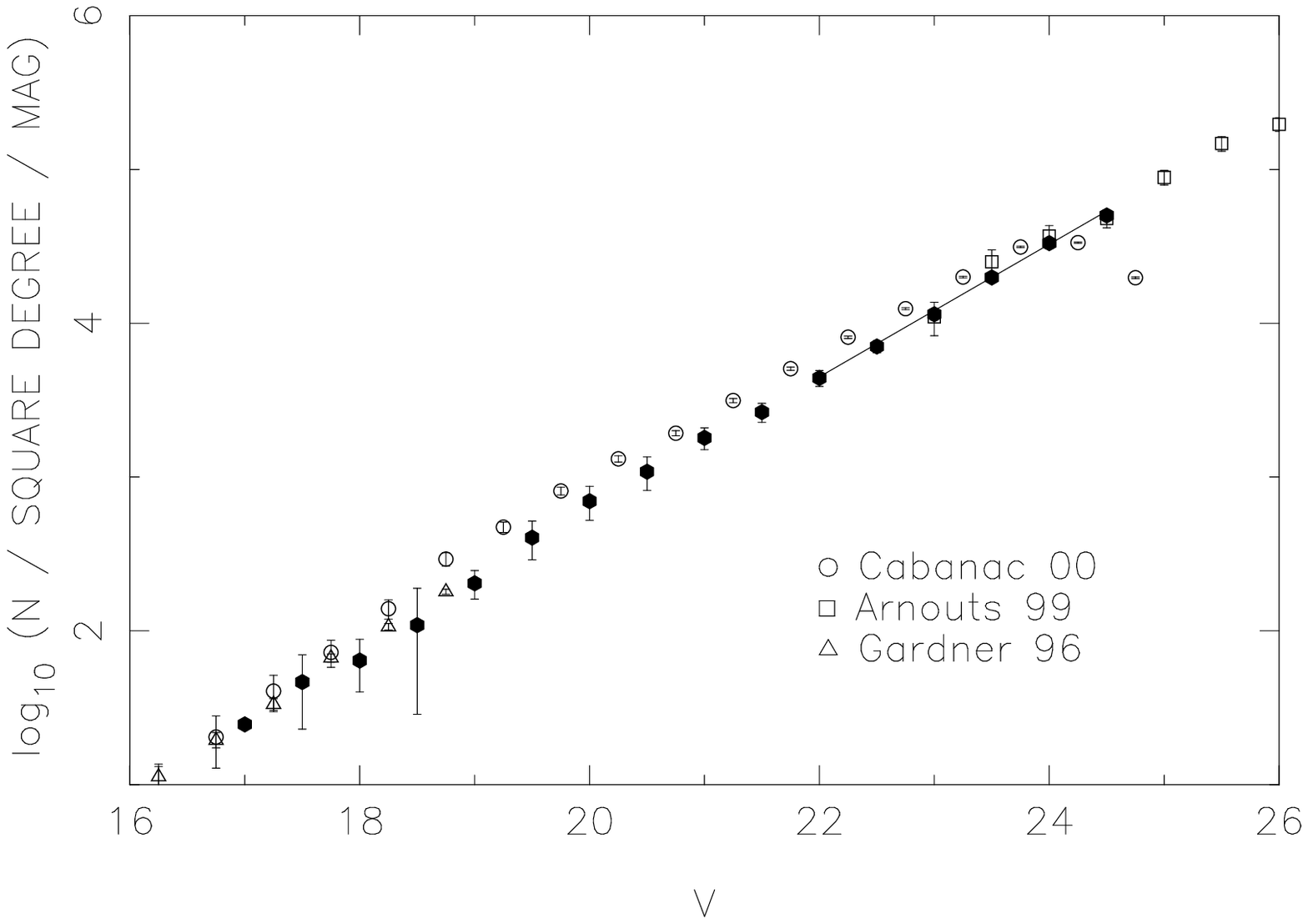,width=\figwidth}
\caption[countsv_all.ps]{
Same as Fig.~\ref{fig:countsi_all} but for $V$ counts.
See Table~\ref{tab:countsv} for values.
The line shows the best fit to our counts 
in the range $22.0<V<24.5$ (slope of $0.43\pm0.02$).
\label{fig:countsv_all}
}
\end{figure}

Figure~\ref{fig:countsi_all} compares our $I$ number counts with those measured
by other groups. Our data are shown by the filled hexagons. The
measured values 
may be found in Table~\ref{tab:countsi}.
The uncertainties at each magnitude 
were calculated from the pointing-to-pointing variations.
Also shown in the figure are the counts of \citet{post-98,arnn-99,cabanac-00,met-01};
and \citet{mcc-01}. A correction of $I = I_{\rm AB} - 0.43$
was applied to the counts of \citeauthor{mcc-01} to bring them to the Cousins 
$I$-band system. Clearly, the counts are in reasonably good agreement
over the whole range.  The highest and lowest values at each
magnitude were obtained by \citeauthor{cabanac-00} and \citeauthor{post-98}
respectively. However, as discussed by \citeauthor{cabanac-00}, that group
systematically measure $20-30\%$ higher counts than
\citeauthor{post-98} but have large (0.2-0.5) 
errors in their zero-point calibration.

The line in Figure~\ref{fig:countsi_all} shows the best fit (Table ~\ref{tab:bestfits}) 
to our counts in the range $21.0<I<23.5$ (slope of $0.31\pm0.01$).
A comparison between our best-fit slope and the values found by 
other authors is shown in Table~\ref{tab:slopes}. Although the range over
which the slope is fit varies slightly, the values are generally in good 
agreement.

Figure~\ref{fig:countsv_all} shows a similar comparison, but for the
(Johnson) $V$ counts (Note that $V_{\rm AB} =$ Johnson $V$).
The filled hexagons in Figure~\ref{fig:countsv_all} represent our
measurements. The values are given in Table~\ref{tab:countsv}. 
$V$-band galaxy number counts are more rarely published in the literature.
We compare to the counts of \citet{gard-96,arnn-99};
and \citet{cabanac-00}. We find that we are in good agreement with
\citeauthor{gard-96} at bright magnitudes and   \citeauthor{arnn-99}
at faint magnitudes. However, the agreement with \citeauthor{cabanac-00}
is rather poor.  \citeauthor{cabanac-00}, however, also report large
uncertainties in their $V$-band calibration (ranging from 0.2 at bright
magnitudes to 0.5 at $V$ of 24).

The line in Figure~\ref{fig:countsv_all} shows the best fit to our
data in the range $22.0<V<24.5$
(slope of $0.43\pm0.02$).
Within uncertainties, this value agrees well with the slopes reported by
\citet{wf-97,driv-94} and \citet{smailhyc-95} (See Table~\ref{tab:slopes}). Our data do not reach a sufficiently faint limit 
to detect any flattening of the slope reported  at $V\sim24.5$
by 
\citeauthor{smailhyc-95}


\section{THE ANGULAR CORRELATION FUNCTION}
\label{sec:wt}

The two-point angular correlation function, \wt, measures the excess
probability (over a random Poisson distribution) that two galaxies
will be found in the solid angle elements $\,d \Omega_{1}$ 
and $\,d  \Omega_{2}$,
separated by angle $\theta$. It is
defined by 

\begin{equation}
\label{eq:prob}
\,d P  = \bar{N}^{2}[1 + \wt]  \,d \Omega_{1} \,d  \Omega_{2}
\end{equation}
where $\,d P$ is the joint probability of finding galaxies in the
two solid angle elements, and $\bar{N}$ is the mean surface
density of galaxies.

\subsection{The Dependence of \wt\ on Apparent Magnitude and Passband}
\label{ssec:2ptfilter}


We compute \wt\ using the \citet{ls-93} estimator.

\begin{equation}
\label{eq:wtls}
\omega (\theta) = \frac{DD - 2DR + RR}{RR}
\end{equation}
where $DD$, $DR$ and $RR$ are the number of data-data, data-random, and random-random pairs (scaled appropriately by the number of data and random points) at angular separations
$\theta \pm \,d \theta$ respectively.
This estimator is based on the N-point function \citep{szap-98} and has been shown 
to have the advantage of reduced edge effects and smallest possible variance.

To determine DR and RR we generated a catalog for each pointing 
containing 50000 random points covering a similar area to the data. We 
masked this random catalog with the same
masks we used to mask out saturated stars and sub-quality regions of the CCDs. 
The remaining catalogs contained $\sim 30000$
randomly distributed points.  

We also applied an integral constraint (IC) correction \citep{grothp-77}.
Estimating the mean galaxy density and the two-point correlation
function from any survey
limited in area results in a \wt\ artificially reduced
by the amount

\begin{equation}
\label{eq:ic}
C = \frac{1}{\Omega^{2}} \int \int \wt  \,d \Omega_{1} \,d \Omega_{2}
\end{equation}
where the integrals are performed over the total solid angle, $\Omega$,
subtended by each of the pointings after masking by the detection
masks \citep{re-99}. The  correction is a function of the effective survey 
area and, to a lesser extent, the survey geometry. 
Because the mean density measured from a large CCD more closely
approximates that of the mean global value, the  correction 
is smaller for larger fields of view. 


In practice \wt\  is calculated
without the correction, equation~(\ref{eq:ic}) is integrated over all
elements of solid angle $\,d \Omega_{i}$ in the survey area
and then \wt\  is recalculated with the correction added.
A stable solution is reached by iteration.
In order to calculate the correction it is therefore
necessary to assume a functional
form for the two-point correlation function.
For local, bright, optically selected galaxy samples 
\wt\ has been shown
to be well approximated by a power law

\begin{equation}
\label{eq:powlaw}
\wt = A_w\theta^{-\delta}
\end{equation}
with $\delta =0.8$, and where \Aw\ is the amplitude of \wt\ (Measuring angular separation in
arcmin as we do here, \Aw\ is then defined as the amplitude of \wt\ at $\theta = 1\amin$).

For each of our pointings, the IC correction, $C$,  was found to be $\sim0.162\Aw$, 
varying slightly
depending on the field geometry and detection mask. The values of $C$
determined for each of the six pointings were 
comparable since the field
sizes and geometries were very similar. Note that the correction only becomes important
when measuring the correlation function for galaxies at large
separation; for the fields-of-view analyzed here, the correction has
a very small effect on \Aw. 

To calculate an error estimate we utilize
the fact that we have six separate pointings and compute the uncertainties
from the field-to-field variations at each
separation. 
This method of determining the uncertainties is superior to most 
other analyses which are limited
to one or two pointings. These are forced to employ  bootstrap resampling
techniques in order to estimate uncertainties.
However, in this analysis, correlations between measurements
on different scales are not taken into consideration, which may 
result in an underestimate of the $\chi^{2}$.

\begin{figure}
\centering\epsfig{file=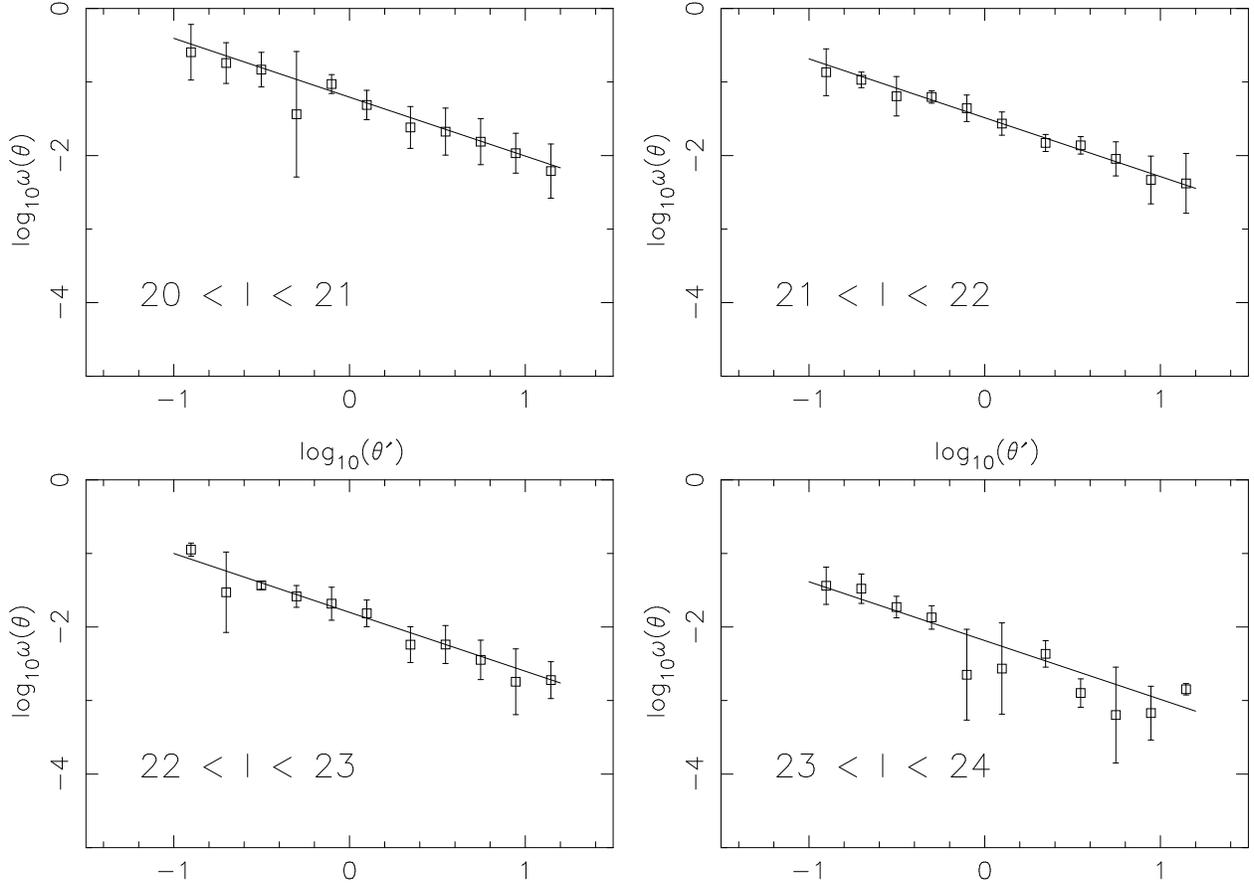,width=\figwidth}
\caption[logwi.ps]{
The logarithm of the angular correlation function, \wt, as a function of the
logarithm of the angular separation in arcmin for 
various $I$-band slices as shown.
The uncertainties are calculated from the pointing-to-pointing variations.
The lines show the best fits (Table~\ref{tab:wt}) in the range $-1<\log_{10} (\theta)<1$ $(6\asec<\theta<10\amin)$ assuming  
a power-law slope of $-0.8$ for \wt. 
 \label{fig:logwi}
}
\end{figure}

 
Figure~\ref{fig:logwi} shows the two-point angular correlation function
for four $I$-band slices, each one-magnitude wide, in the range
$20.0<I\leq24.0$. 
We calculated \wt\ using  logarithmically
spaced bins of width $\Delta \log \theta = 0.2$. 
On small scales ($\log_{10}(\theta) < 0.2$, $\theta < 1.58 \amin$) we
estimated 
the two-point function
(equation~\ref{eq:wtls}) directly from pair counts.
On larger scales, \wt\ was determined from counts in cells. 
The four lines in Figure~\ref{fig:logwi} show the
best fit to the data in the range  $-1<\log_{10}(\theta)<1$ 
($6\asec < \theta < 10 \amin$ ),  assuming $\delta = 0.8 $. (We 
also fitted \Aw\ and the slope $\delta$ separately, both with and
without an IC correction, and found 
$\delta =0.8$ to be a good fit to the correlation function for
each of the four magnitude intervals).
The best-fit value of $\log_{10}\Aw$, the logarithm of the amplitude
of \wt\ at $1\amin$, as a function of median $I$ magnitude is
shown in Table~\ref{tab:wt}. We find a monotonic decline in \Aw\ with
increasingly faint median $I$ magnitude.

\begin{figure}
\centering\epsfig{file=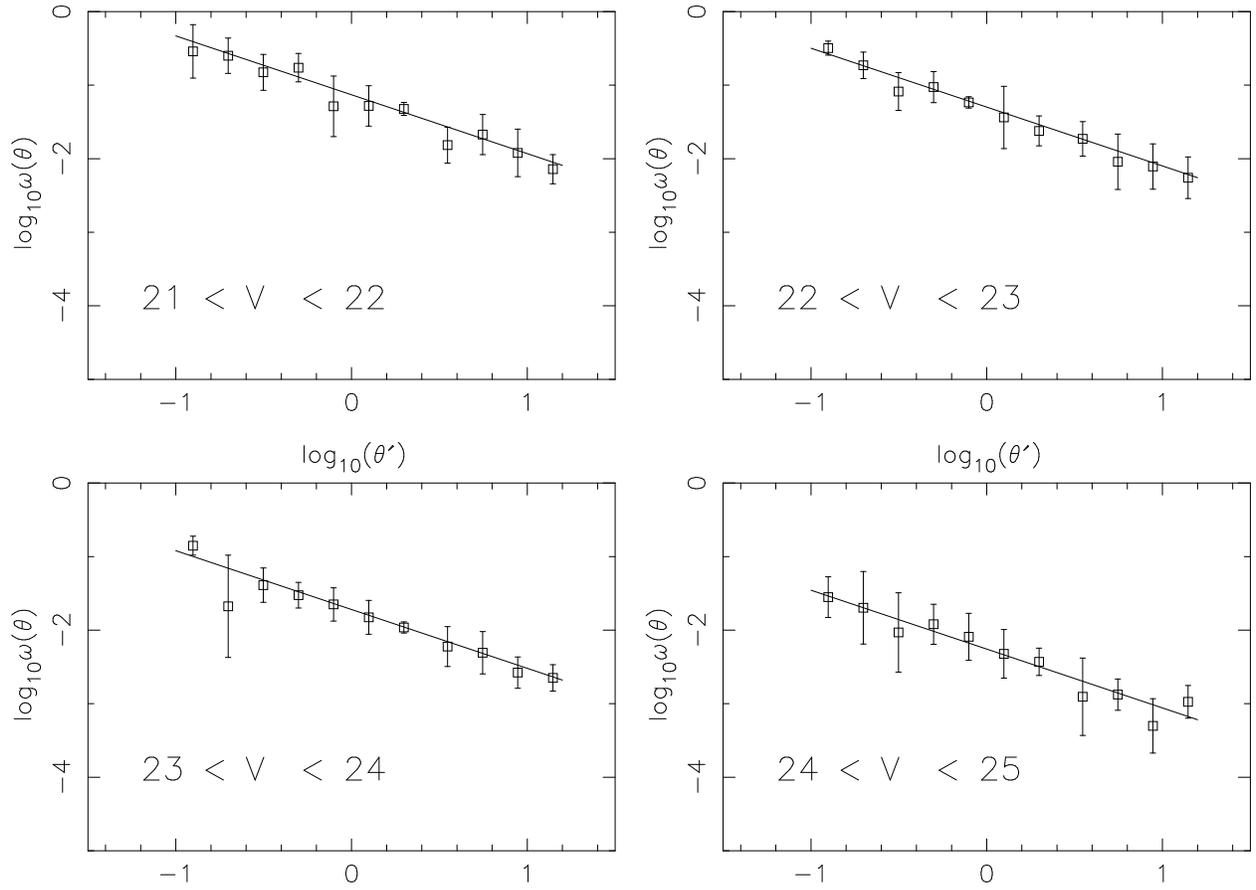,width=\figwidth}
\caption[logwv.ps]{
As for Figure~\ref{fig:logwi} but for $V$-passband. 
 \label{fig:logwv}
}
\end{figure}

Figure~\ref{fig:logwv} shows our measurements of \wt\, but for the
$V$-band, for four one-magnitude wide slices in the range $21.0<V\leq25.0$.
As with the $I$-band, we find a monotonic decline in the amplitude of the
two-point correlation function with increasingly faint $V$ magnitude.
The best-fit values of $\log_{10}$\Aw\ as a function of median $V$ magnitude  
may also be found in Table~\ref{tab:wt}. 
 
\begin{figure}
\centering\epsfig{file=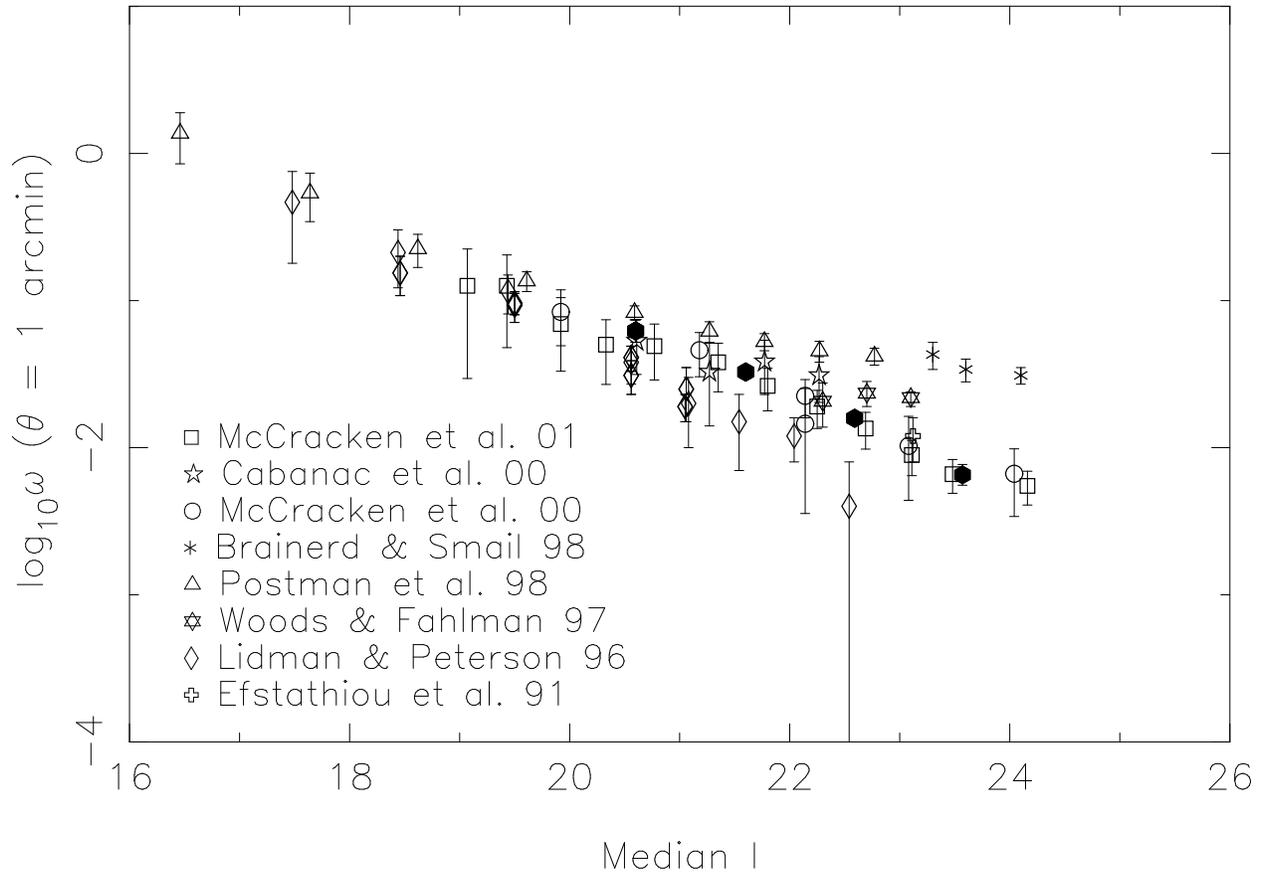,width=\figwidth}
\caption[ampi.ps]{
The logarithm of the amplitude of the angular correlation function
\wt\ at 1$\amin$ for the $I$-passband plotted as a function
of median magnitude (see Table~\ref{tab:wt}). 
The filled hexagons represent our data.
Also shown are the measurements from other groups.
 \label{fig:ampi}
}
\end{figure}

Figure~\ref{fig:ampi} compares our measured values of clustering amplitude at
$1$ arcmin (as a function
of median Cousins $I$ magnitude)
with the values obtained by other groups. We compare with the results
of \citet{efs-91, lp-96, wf-97, bs-98, post-98, cabanac-00, mcc-00} and
\citet{mcc-01}. 
At bright magnitudes ($I<20$) there is good agreement
among the measured values of \Aw.
However, at fainter magnitudes some discrepancies appear.
Our measurements are in very good agreement
with those of \citeauthor{efs-91} (using the value quoted in
\citeauthor{lp-96} Table 6), \citet{mcc-00} and \citet{mcc-01}. 
The clustering amplitude we measure is stronger than that 
found by \citeauthor{lp-96}. We 
measure a somewhat lower clustering amplitude
than \citeauthor{wf-97} and \citeauthor{cabanac-00}.  We measure  a
significantly lower clustering amplitude 
than  \citeauthor{bs-98} and \citeauthor{post-98}

The discrepancy between our results and those of \citeauthor{wf-97}
and \citeauthor{bs-98}
may be explained by the relatively small area investigated by those studies, 
and the corresponding inherently large uncertainties. 
\citeauthor{wf-97} had field sizes 
of $3 \times 49$ arcmin$^{2}$ and IC correction $\sim0.02$. The corresponding
values for 
\citeauthor{bs-98} were $2 \times 30$ arcmin$^{2}$ and IC $\sim0.02$.
\citeauthor{cabanac-00} report large uncertainties in their
photometric zero points which increase at fainter magnitudes.
Any variations in effective depth due to zero-point errors
or  variations in observing condition (\eg\ variations in the seeing) between
exposures would artificially mimic large-scale power. 
The discrepancy with the results of the survey
of \citeauthor{post-98} is more puzzling. The large survey of 
\citeauthor{post-98} consisted of 16 $\deg^{2}$ ($256 \times 16$ arcmin$^{2}$, 
$IC\sim0.002$).
They find a 
shallower best-fit slope to the correlation
function ($\delta  \simeq0.7$) for $I>22$. 
This disagreement notwithstanding, it seems likely that
zero point variations from frame-to-frame may remain in the Postman survey. 

\begin{figure}
\centering\epsfig{file=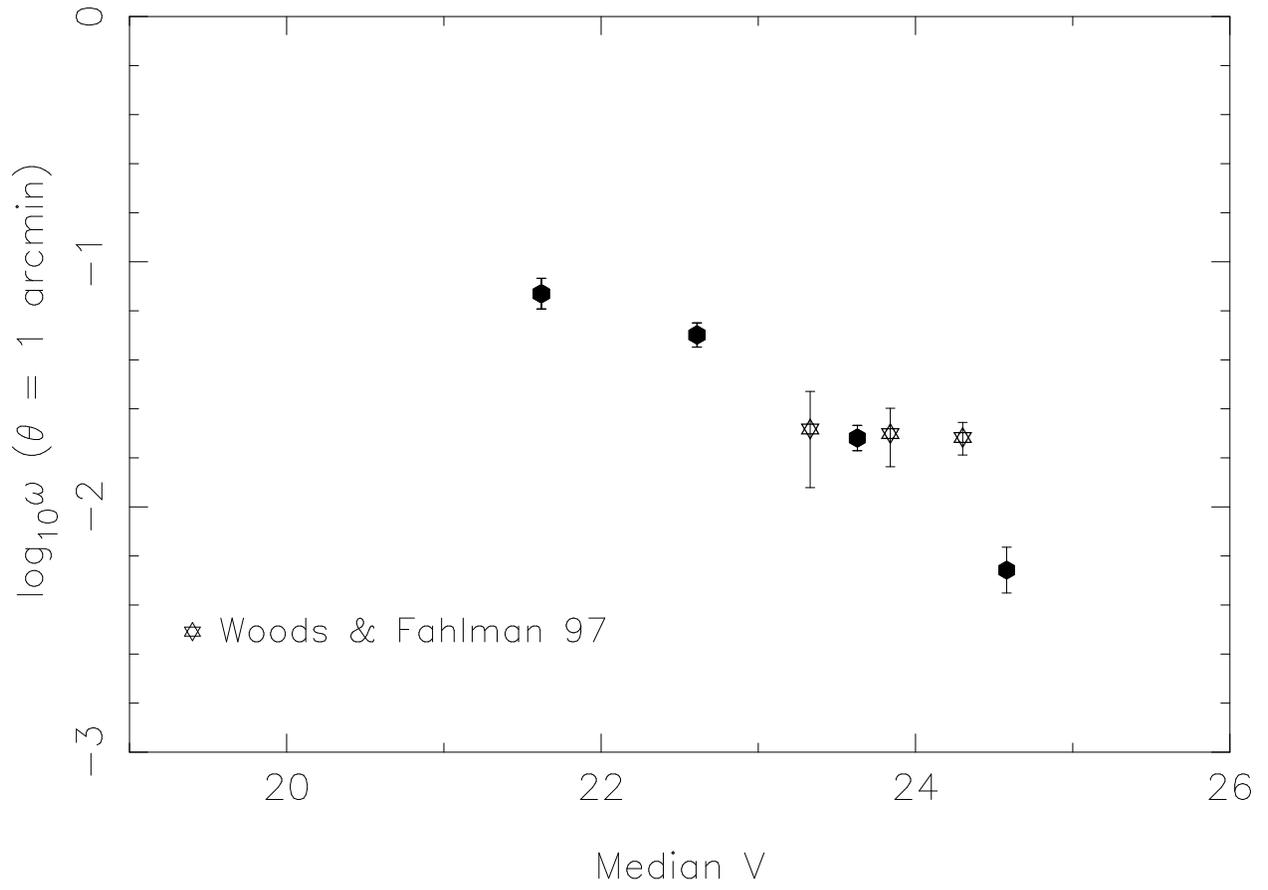,width=\figwidth}
\caption[ampi.ps]{
Same as Fig.~\ref{fig:ampi} but for $V$-passband.
\label{fig:ampv}
}
\end{figure}

Figure~\ref{fig:ampv} compares our measured values of clustering amplitude at $1$ arcmin in the $V$-band
with the values obtained by \cite{wf-97} (the only other 
measurements of \Aw\ published in this passband). The uncertainties  associated 
with the 
\citeauthor{wf-97} measurements are larger than for our measurements
but the agreement is generally good at $V < 24$. 
We measure a significantly (factor of two) lower clustering 
amplitude than the \citeauthor{wf-97}
faintest determination (at $V=24.3$).

In the next section we predict the correlation function amplitude
that we would expect to measure as a function of apparent magnitude, based on deep spectroscopic observations.
We then compare these predictions to our 
measurements.
We investigate whether our observations can be matched to the predictions 
assuming simple analytic models for galaxy clustering evolution.

\subsection{Clustering Strength Predictions}
\label{ssec:invert}

As mentioned in \S~\ref{sec:intro} the two-point spatial correlation function, \xr, for local, optically
selected galaxies has been shown to be well described 
by a power law 

\begin{equation}
\label{eq:xr}
\xr = {\left( {r}/{\rnought} \right)}^{-\gamma}
\end{equation}
where r is physical (proper) separation and $\rnought$ is the correlation length at $z=0$.
Robust determinations for the power-law slope of $\gamma \simeq 1.8$ and
physical correlation length of $\rnought \simeq 5.0 \hMpc$ have been made
by many groups
\citep{dp-83,love-95, post-98, willmer-98, carl-00, norb-01}. 

\citet{grothp-77} proposed the ``$\epsilon$-model'' to describe
the evolution with redshift of the correlation function, measured
in terms of proper separation:

\begin{equation}
\label{eq:xrz}
\xrz = {\left( \frac{r}{\rnought} \right)}^{-\gamma} (1+z)^{-(3+\epsilon)}
\end{equation}
There are several noteworthy values of the evolutionary parameter $\epsilon$.
The clustering pattern is fixed in comoving coordinates if
$\epsilon = -1.2$ ($\epsilon = \gamma -3$): galaxy clusters expand with the
universe and clustering does not grow with time.
If $\epsilon = 0.0$,
``stable clustering'', 
then clustering is fixed in proper coordinates. In this case, the galaxies
are dynamically bound and stable at small
scales. The clustering grows in this case because the background
density of galaxies is diluted by the expansion: it is effectively the voids that
are growing.
Linear theory predicts $\epsilon = 0.8$ to
describe the evolution in the clustering pattern of \emph{dark matter} 
in an $\omegam=1$ universe \citep{baugh-99} (See also \eg\
\citet{colin-97, carl-00} and \citet{kauffhiz-99} for $\epsilon$ predictions 
for the dark matter evolution in alternative cosmologies).
If $\epsilon > 0$ then clustering grows with time
in proper coordinates, as expected from gravitational collapse.

The relation then (for small angles) between the two-point angular and spatial coordinates, \wt\ and \xrz\, is given by Limber's equation \citep{limb-53,peeb-80}

\begin{equation}
\label{eq:wtrnought}
 \wt = \frac{\Gamma(1/2)\Gamma[(\gamma-1)/2]}{\Gamma[(\gamma/2)]} \frac{A}{\theta^{\gamma-1}} \rnought^{\gamma}
\end{equation}
where the Gamma function factor 
equals 3.68 for $\gamma=1.8$. $A$, the
amplitude factor,  is given
by,

\begin{equation}
\label{eq:A}
 A = \frac{\int_{0}^{\infty} g(z) {\left [ \frac{dN(z)}{dz} \right ]
 }^{2}\,dz}{  {\left[ \int_{0}^{\infty}  \frac{dN(z)}{dz} \,dz\right]}^{2}}
\end{equation}

Here, $dN/dz$ is the number of
galaxies per unit redshift interval. The function g(z) depends only on
$\gamma$, $\epsilon$, and the cosmology 
\begin{equation}
\label{eq:g}
g(z) = \left ( \frac{dz}{dx} \right ) x^{1-\gamma}F(x)(1+z)^{-(3+\epsilon-\gamma)}
\end{equation} 
$F$ gives the correction for curvature,
\begin{equation}
\label{eq:F}
F(x)^2 = 1 + \omegak{\left( \frac{\Hnought x}{c} \right) }^2 
\end{equation}
$x$ is the comoving distance at redshift z,
\begin{equation}
\label{eq:comoving}
x = \frac{c}{\Hnought} \int_{0}^{z}   \frac{1}{E(z)}
\end{equation}
$E$ is  given by
\begin{equation}
\label{eq:E}
E  =  \left[ \omegam (1+z)^3 + \omegak (1+z)^2 + \omegal \right] ^{\onehalf}
\end{equation}
and \omegak\ (the ``curvature of space'') is defined by 
$\omegam + \omegal + \omegak =1 $.
\citep{cpt-92, hogg-99, brown-00} 


Therefore, predictions of galaxy clustering strength, \wt\, are dependent on a
number of factors. Namely, these are $\gamma$, $\rnought$, $\epsilon$, the
assumed redshift distribution $N(z)$, the Hubble constant $\Hnought$, and
one's choice of cosmology.
If the spatial two-point correlation function is well described by a power
law (equation~\ref{eq:xr}),
then the angular two-point correlation function 
will also be well described by a power law of slope $\delta = \gamma - 1$
(equation~\ref{eq:wtrnought}). The physical correlation length at $z=0$, 
\rnought\ (to the power $\gamma$), 
determines the normalization of \wt. In making our predictions we
adopt a local correlation length  of $\rnought=4.9 \hMpc$, the 
best-fit  real space \rnought\ determined by the
2dFGRS \citep{norb-01}. 

Accurate predictions of the correlation amplitude at each magnitude
depend crucially on the assumed redshift
distribution (Note that \wt\ decreases with increasing apparent
magnitude because of the increasing probability of chance projected alignments of
galaxies at very different redshifts). \wt\ is extremely sensitive to the
shape of the assumed redshift distribution but not to
its normalization.
As in \citet{wklc-01}, in making our predictions, we 
adopt redshift distributions appropriate to each magnitude interval
based on spectra from the SSA22 field sample of Cowie
\citep{cghskw-94,cshc-96, csb-99, wcbb-02}.
We model the normalized redshift distribution 
as
\begin{equation}
\label{eq:pz}
p(z)  = 0.5 z^{2} \exp(-z/\znought) / \znought^{3}
\end{equation}
for which the mean redshift is $\overline{z} = 3 z_0$ and the
median redshift is $z_{\rm median} = 2.67 z_0$.
Note that equation (\ref{eq:pz}) has only one free parameter,  the
redshift scale parameter $z_0$.

We calculate $z_0$ (in half-magnitude intervals) by 
setting the mean redshift 
to match that from the Cowie sample (allowing for 
varying numbers of galaxies  in each magnitude interval).
The Cowie sample reaches a limiting magnitude of $I \simeq 23.5$, $V \simeq 24.5$.
At the faintest magnitudes the sample becomes about $20\%$ incomplete.
It is thought that the galaxies for which a redshift cannot
be determined lie predominantly around $z=1.5-2.0$. 
We therefore calculate $z_0$ 
in two ways. Firstly we utilize only
galaxies in the Cowie sample with secure redshifts, our
``raw'' model. Secondly we assign a redshift of 1.8 to the
galaxies from the Cowie catalog without secure redshifts, our
``corrected'' model.
This increases the value of the redshift scale parameter $z_0$ 
we employ, only at the faintest magnitudes where incompleteness 
becomes important. 
It is likely that the true redshift distribution lies between
these two extremes. The value of $z_0$ we adopt to describe the
redshift distribution in each 
half-magnitude interval is shown in Table~\ref{tab:znought}.

In making these predictions of \Aw, 
we assume a flat lambda 
($\Omega_{{\rm m}0} = 0.3, \Omega_{\lambda 0} = 0.7$) 
cosmology with $\Hnought = 100 $ $h\kmsMpc$. The predictions are rather insensitive to the
choice of cosmology, because the majority of galaxies lie
at relatively low  redshift, especially at
bright magnitudes. Adopting an Einstein de-Sitter
cosmology would decrease the predictions by about $0.2$ in the
 log at the faint end.

\begin{figure}
\centering\epsfig{file=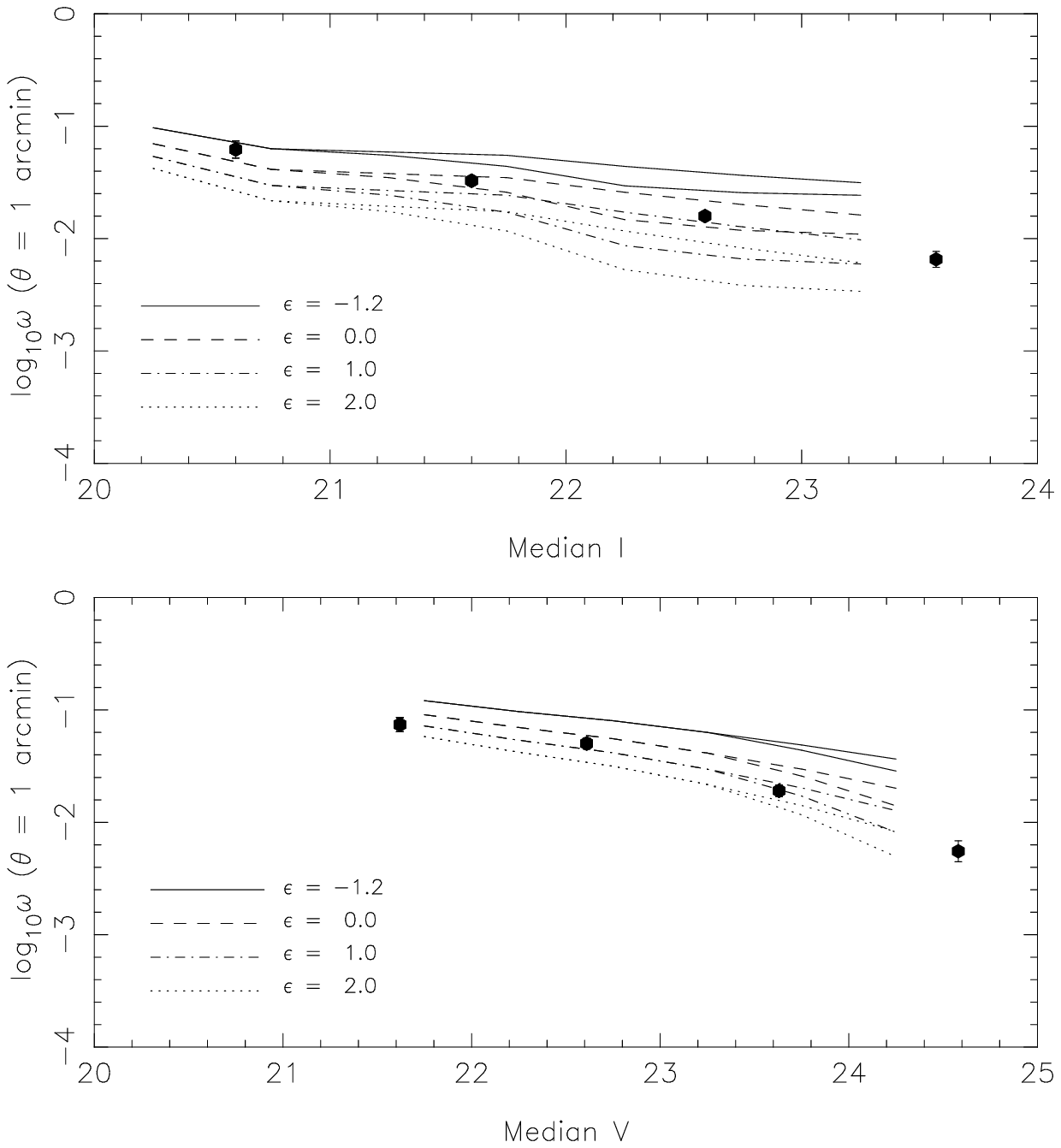,width=\figwidth}
\caption[pred.ps]{
Upper panel shows data (solid hexagons) as in Figure~\ref{fig:ampi} 
for the $I$-band. The lines show our predictions assuming redshift distributions based on
deep spectroscopic observations,  correlation length $\rnought=4.9\hMpc$ as measured
by the 2dFGRS,
and evolutionary parameter $\epsilon = -1.2$ (solid), $0.0$ (dashed),
$1.0$ (dot-dashed) and $2.0$ (dotted). 
Clearly, equation~(\ref{eq:xrz})
does not provide a satisfactory fit to the data for any value of $\epsilon$.
Lower panel shows same but for the $V$-band  data as in Figure~\ref{fig:ampv} (see Table~\ref{tab:wt}
for values).
 \label{fig:pred}
}
\end{figure}

The upper panel of  Figure~\ref{fig:pred} shows 
the $I$-band clustering amplitudes from  Figure~\ref{fig:ampi} (for clarity
we do not plot the values obtained by other groups). 
Also shown are the predictions for \Aw\
assuming redshift distributions based on the Cowie sample
and also assuming an evolutionary parameter 
of either $\epsilon = -1.2$ (solid), $0.0$ (dashed),
$1.0$ (dot-dashed) or $2.0$ (dotted). The upper line in each case
shows the predictions assuming the raw redshift distribution, the
lower line the incompleteness-corrected distribution.
Clearly, equation~(\ref{eq:xrz})
does not provide a very satisfactory fit to the data 
for any  value of $\epsilon$, assuming a smooth extrapolation of
the clustering evolutionary model out to this study's fainter
magnitude limits. 
We find that a
clustering pattern fixed in physical coordinates, $\epsilon=0$, provides a
reasonable fit to the data at the bright end, but that the data
fall below the predictions at the faint end. 

The lower panel of Figure~\ref{fig:pred} shows the predictions for
\Aw\ for our 
$V$-band data (see Table~\ref{tab:wt} for measured values of \Aw\
 and Table~\ref{tab:znought} for the 
$\znought$ adopted to describe the redshift distribution of each magnitude interval).
As for the $I$-band, we conclude that equation~(\ref{eq:xrz})
does not provide a very satisfactory fit to the data for any value of $\epsilon$ (assuming a smooth extrapolation of
the clustering evolutionary model out to this study's fainter
magnitude limits).

In calculating the predicted correlation amplitude
we utilized a correlation length of
$\rnought=4.9 \hMpc$ as measured by the 2dFGRS. Adopting a larger or smaller local correlation 
length would shift the predictions vertically upwards or downwards.

We conclude that equation~(\ref{eq:xrz}) does not provide a
good fit to the observed evolution in clustering amplitude with redshift.
This conclusion was also reached by \cite{mcc-01}.
We interpret the failure of the predictions to match the data 
as being due to the assumptions of the model being overly simplistic. 
Equation~(\ref{eq:xrz}) implicitly assumes that galaxies with 
different morphologies have similar intrinsic clustering properties. 
This is known not to be the case locally.
\cite{dg-76, ghc-86, mauro-91, iov-93, love-95, hermit-96,guzzo-97,willmer-98} and \cite{norb-02}  have found early-type
galaxies to cluster more strongly than late types.
The morphological
mix of galaxies in any sample will
vary with apparent magnitude due to different
K-corrections for early and late type systems. As one probes 
fainter in apparent magnitude, a higher preponderance of late types
enter the sample \citep{maglio-00}.
Moreover, by selecting on increasingly faint apparent magnitude one samples
intrinsically fainter and likely more weakly clustered galaxies.
\citep{mcc-01}.


We further investigate the bivariate dependence of galaxy clustering evolution
on luminosity and  morphology in the remainder of this paper.
In the next section we investigate \wt\ as a function of $V-I$ color.

\subsection{The Dependence of \wt\ on $V-I$ Color}
\label{ssec:2ptcolor}

In this section we investigate the dependence of
clustering strength on galaxy $V-I$ color.
We employ the same catalogs as used in \cite{wkl-01}, containing 
galaxies which have been detected in \emph{both} $I$ and $V$ images
above a  threshold significance of $4\nu$ (to ensure 
that any given ``detection'' is truly a real object).
Galaxies tend
to be detected at higher significance in the $I$-band images. 
We subdivided the data into seven 
intervals with $V-I$ color ranging from 
0.0 to 3.5 and $\Delta(V-I) = 0.5$ (Table~\ref{tab:wt_iv} shows the
number of galaxies in each color interval). 
As in section~\ref{ssec:2ptfilter} we
again assume that \wt\ is well described by a  
power law of slope $\delta =0.8$ and we fit
over the range $-1.0 < \log_{10}\wt \leq 1.0$.

\begin{figure}
\centering\epsfig{file=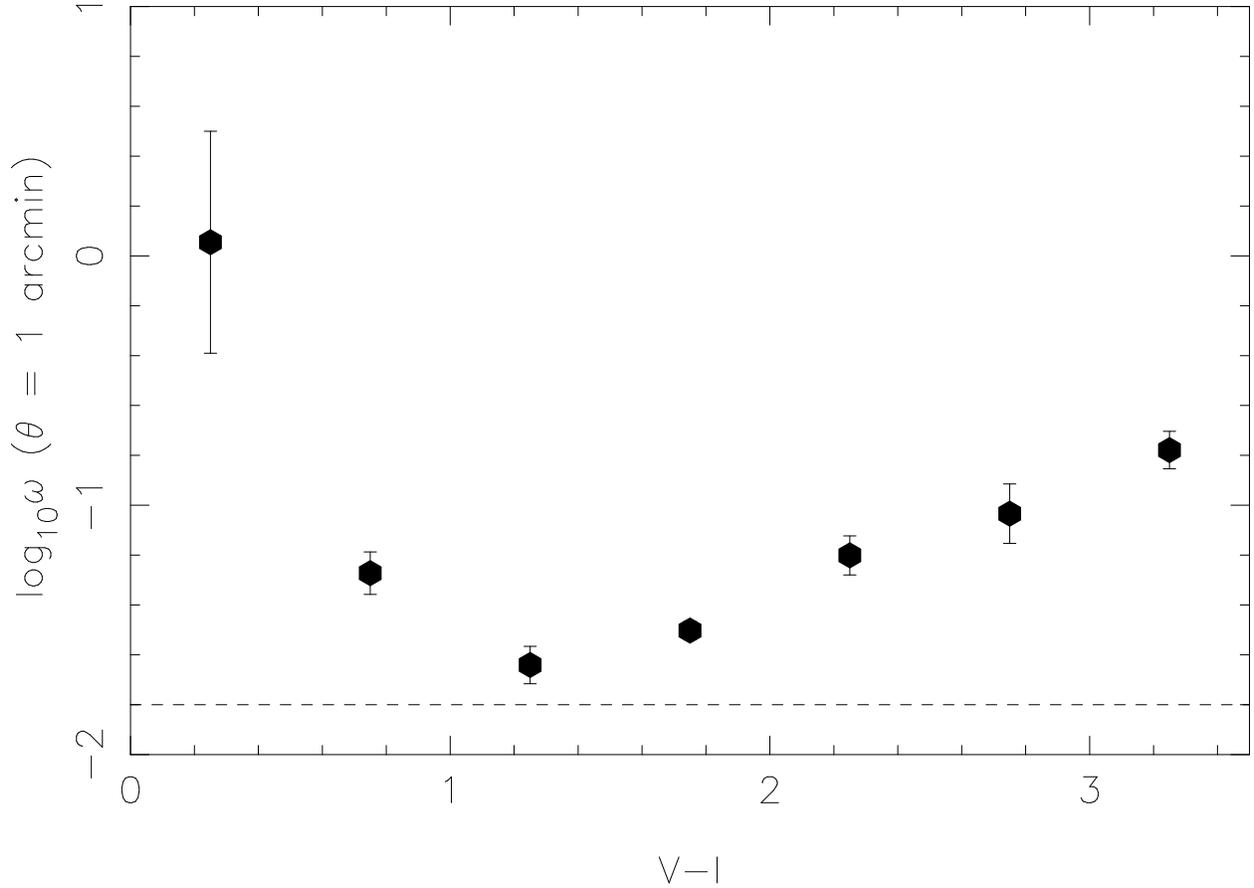,width=\figwidth}
\caption[ampiv.ps]{
The logarithm of the amplitude of the angular correlation function
\wt\ at 1$\amin$ as a function of $V-I$ color for 
galaxies in the magnitude range $20<I\leq23$
(see Table~\ref{tab:wt_iv}). Extremely red and extremely blue galaxies are seen 
to cluster more strongly than those of
intermediate color. The dashed line shows the amplitude measured 
in \S~\ref{ssec:2ptfilter} for the full field sample with median
$I$ magnitude of 22.59. In all cases, the clustering amplitude is
higher than for the full field population.

 \label{fig:ampiv}
}
\end{figure}


Figure~\ref{fig:ampiv} shows  clustering amplitude as
a function of color for galaxies in the range $20.0 < I \leq 23.0$.
Figure~\ref{fig:countsv} shows a turnover in the $V$-band counts
at $V\simeq 25$ and thus we may be subject to missing fainter galaxies 
from the redder ($V-I >2$) intervals. Table~\ref{tab:wt_iv} shows the best-fit values
of \Aw\ obtained for each color interval. As in \S~\ref{ssec:2ptfilter}, the uncertainties
were estimated from the field-to-field variations between our six pointings.
The dashed line in Figure~\ref{fig:ampiv} shows the correlation
amplitude  ($-1.80$ for a median $I$ magnitude of 22.59) obtained for the full field sample in
\S~\ref{ssec:2ptfilter}
(Table~\ref{tab:wt}).

As seen from  Figure~\ref{fig:ampiv},
red galaxies ($V-I\simeq3$) have a clustering amplitude
about $10$ times larger than that measured  for the full
sample in \S~\ref{ssec:2ptfilter}. 
As we shall discuss further
in \S~\ref{ssec:2ptE0}, galaxies with color $V-I=3$ 
are exclusively early types occupying a
narrow range in redshift at about $z=1$. 
Since the redshift distribution
N(z) for these red galaxies is rather narrow, chance alignments of
galaxies of the same color at significantly higher or
lower redshift cannot occur and hence
the measured angular correlation function is not diluted
as it is for galaxies with colors in the range $1< (V-I)< 2$,
which are more widely distributed in redshift.
The high clustering amplitude found for extremely red galaxies 
is also due in small part to these being 
of a morphological type which (at least locally) cluster 
more strongly than later types. 

As noted by \citet{mcc-01}, we also find an upturn in the clustering
amplitude of extremely blue galaxies ($V-I\simeq0.5$). A similar effect was noted by
\citet{lsk-96}. The most
likely explanation for the strong angular clustering amplitude
found for the blue galaxies is that these
galaxies are all situated at  relatively low redshifts. As with the red
sample, the redshift distribution N(z) of these extremely blue galaxies 
 is rather narrow and  thus \Aw\ is not diluted as
strongly in projection over redshift as is the intermediate sample. 

The results of this section and of \S~\ref{ssec:invert} 
strongly suggest that studying and interpreting galaxy clustering
either as a function of 
median magnitude or of color is greatly complicated by the varying 
morphologies and intrinsic luminosities of galaxies contained
in the different samples.
A far more promising approach would be to isolate
the same population and study its clustering evolution with redshift.
Given fluxes in only two passbands, as we have here, it is impossible
to select and
measure evolution in \wt\ for all morphological types.
However, as we show in the next section, it is possible to
use crude photometric redshift determinations
to isolate and analyze the clustering evolution
of \Lstar\ early-type galaxies with redshift.


\subsection{The Dependence of \wt\ on Morphological Type}
\label{ssec:2ptE0}

As in \citet{wklc-01}, we use $V -I$ color
combined with an $I$-band magnitude cut 
to select a sample of bright early-type galaxies.
This technique depends on the fact that
early-type galaxies are the reddest galaxies at any given redshift. 
By selecting galaxies of some color we see a superposition
of early types at redshift $z_E$ such that $c = c_E(z_E)$
and later types at their appropriate, but considerably higher, redshift. 
An $L \sim L_\star$ early-type galaxy 
appears much brighter than an $L \sim L_\star$ spiral galaxy, and
as explained in more detail
in \S 2.2 of \citeauthor{wklc-01}, with
a judicious cut in $I$ magnitude 
it is possible to isolate a bright early-type galaxy sample.

We divide the data into nine ($\Delta z = 0.1$) bins (the lowest
redshift $z=0.1\pm0.05$ 
bin contains very few galaxies so is discarded here).
In addition to a morphological dependence, there has recently been
quite considerable evidence in the literature
for a luminosity dependence: luminous galaxies clustering more strongly than their fainter counterparts.
Such a luminosity dependence has been reported by 
\citet{benoist-96, guzzo-97, willmer-98, small-99, norb-01, firth-02}
and \citet{zehavi-02}.
The evidence for a dramatic increase 
in clustering strength with increasing luminosity for $L > \Lstar$ 
galaxies is the most compelling.
Indeed, \citet{norb-02} find a factor of 2.5 difference between the
clustering strength of \Lstar\ and 4\Lstar\ galaxies and 
suggest that it is in fact luminosity and not type,
which is the dominant factor (\citeauthor{norb-02} measure a
clustering strength for early types
about $50\%$ higher than that for late types at all luminosities).

In an effort to eliminate any complicating effect of 
a luminosity-dependent component to \wt, we
decided to make a further restrictive cut to the early-type galaxies
we allowed into our sample at each redshift. We chose to exclude
all galaxies with absolute magnitude fainter than 
$M = M_* + 1$ from our analysis.
This ensures that the remaining sample has an effective
luminosity $\Leff\sim\Lstar$ ($\Leff=0.98\Lstar$ 
if one assumes the 2dFGRS  value of $\Lstar = -19.61$ 
\citep[See also  \citealt{madg-01})]{folkes-99}. 
This assumes that $\Lstar$ for early-type galaxies 
does not evolve between $z=0$ and 1.
Based on our knowledge about early-type evolution with redshift 
this does not seem a grossly inaccurate assumption \eg\ \citet{lilly-95}
found that the red (redder than 
present-day Sbc and hence early type) 
sample from the Canada-France
redshift survey, was consistent with no change in $\Lstar$ 
between $z\sim0.8$ and $z\sim0.3$ (their red sample was also consistent
with a change of \emph{at most} a few tenths of a magnitude)

Table~\ref{tab:wt_E} shows the number of galaxies in each redshift 
interval which meet our criteria. In estimating \Aw, we
once more assumed that \wt\ was well described by a  
power-law with slope $\delta =0.8$ ($=\gamma-1$).
Slightly steeper slopes have been
claimed for the spatial correlation function slope $\gamma$
for early-type galaxies \ie\ 
$1.87\pm0.07$ \citep{love-95},
$2.05\pm0.1$ \citep{guzzo-97}, 
$1.91\pm0.06$ \citep{willmer-98},
$1.91\pm0.06$ \citep{shep-01},
and $1.87\pm0.09$ \citep{norb-02}.
A slightly steeper slope would have negligible effect
on our conclusions, and is far from being
the main source of uncertainty.
Note also that by measuring the correlation function amplitude
at a fixed angular scale of $1\amin$ we are measuring at
increasing projected radius with redshift, ranging from about $140\hkpc$ at
$z=0.2$ to about $330\hkpc$ at $z=0.9$. This analysis, therefore,
assumes that any bias is not a function of scale \citep{maglio-00}.

\begin{figure}
\centering\epsfig{file=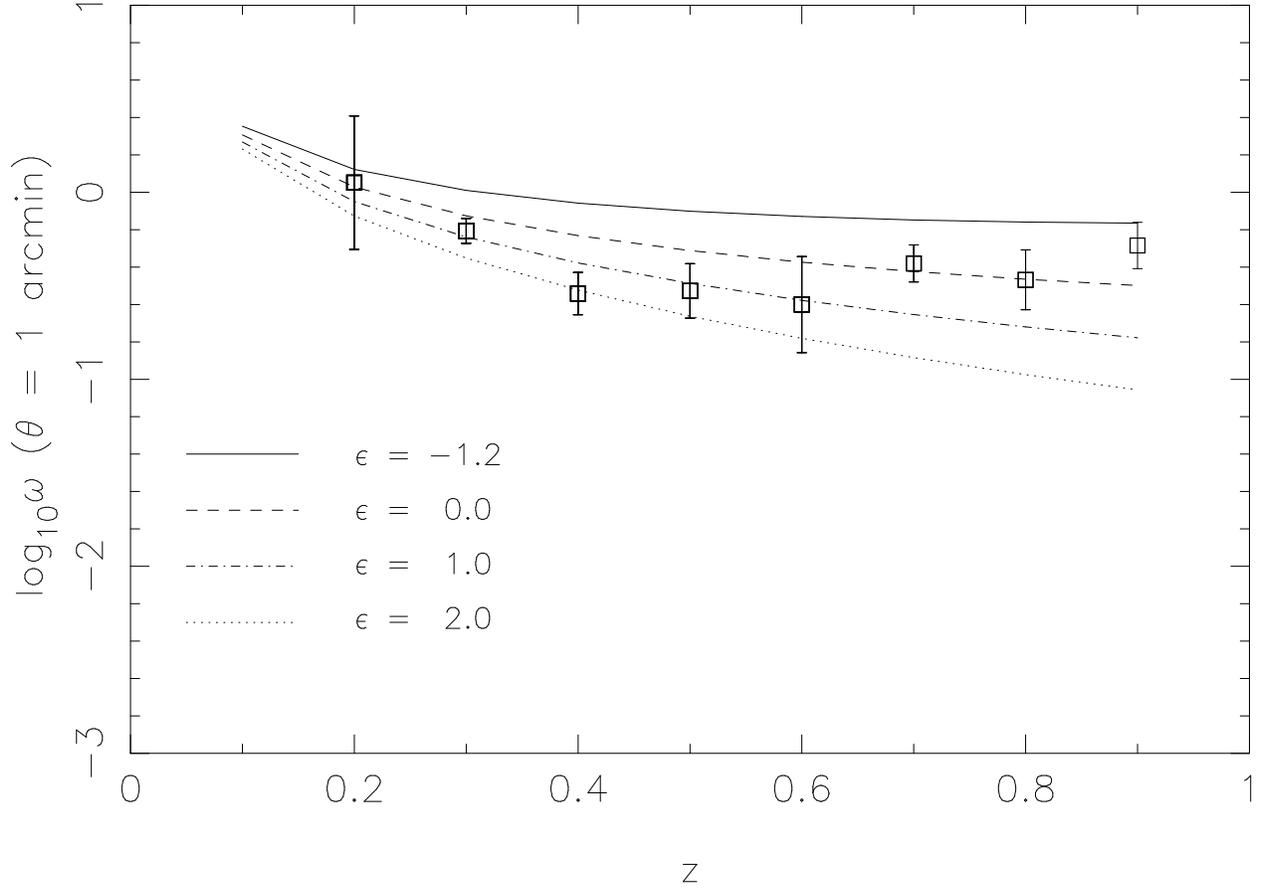,width=\figwidth}
\caption[ampE.ps]{
The logarithm of the amplitude of the angular correlation function
\wt\ at 1$\amin$ for $\Lstar$  early-type galaxies 
as a function of redshift. The lines show  predictions 
for evolutionary parameter $\epsilon = -1.2$ (solid), $0.0$ (dashed),
$1.0$ (dot-dashed), and $2.0$ (dotted),
assuming $\rnought=5.7\hMpc$, the correlation length of early-type galaxies
determined locally by the 2dFGRS.
 \label{fig:ampE}
}
\end{figure}

Figure~\ref{fig:ampE} shows the logarithm of $\Aw$, the amplitude
of \wt\ at $1\amin$,
as a function of redshift for early types. The measured values of the
two-point correlation function amplitude at each redshift may be found
in Table~\ref{tab:wt_E}. For comparison, superimposed on
Figure~\ref{fig:ampE} are the predictions assuming a 
correlation length of $\rnought=5.7\hMpc$, 
the value determined by the 2dFGRS for local
early types with $L=\Lstar\pm0.5$
\citep[their Table 2]{norb-02}. The various lines indicate the predictions
assuming evolutionary parameter 
$\epsilon = -1.2$ (solid), $0.0$ (dashed),
$1.0$ (dot-dashed) or $2.0$ (dotted).
The resulting predictions assuming the 2dFGRS value of
$\rnought$ match our measurements of the clustering
amplitude reasonably well over the whole range of redshift.
Clustering fixed in comoving coordinates, $\epsilon = -1.2$,
appears somewhat to overestimate the measured clustering amplitude.
At the other extreme, rapid growth in clustering with time,
$\epsilon = 2.0$, appears rather to underestimate the actual data.

\begin{figure}
\centering\epsfig{file=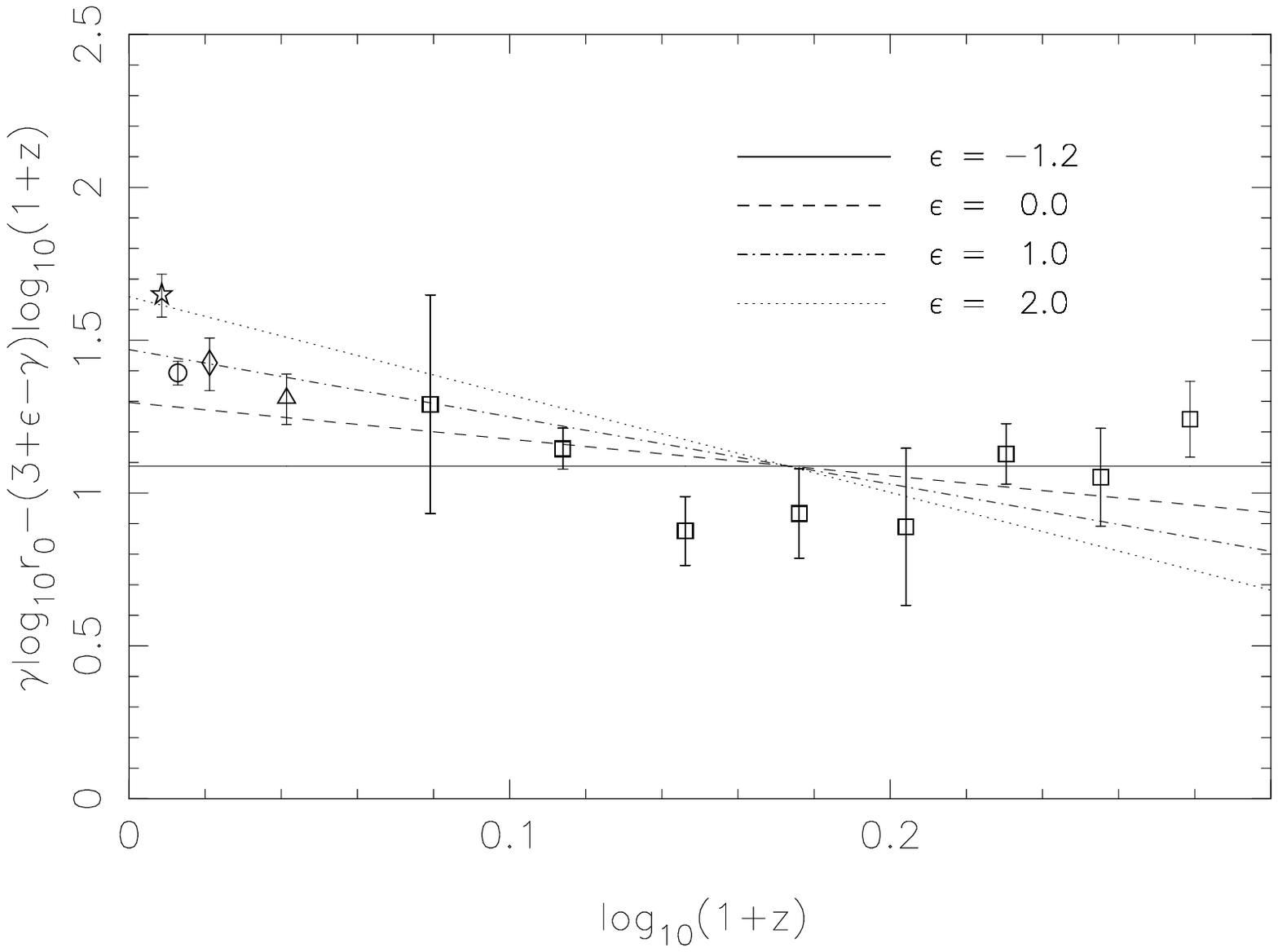,width=\figwidth}
\caption[ampEfancy.ps]{
The logarithm of the cosmology-independent component 
of the angular correlation function
\wt\ at 1$\amin$ for $\Lstar$ early-type galaxies
assuming $\gamma=1.8$. The lines show best-fit
physical correlation length \rnought,
for evolutionary parameters $\epsilon = -1.2$ (solid), $0.0$ (dashed),
$1.0$ (dot-dashed) or $2.0$ (dotted). Also shown are the values
measured by \citeauthor{love-95} (diamond), \citeauthor{guzzo-97} (star), 
\citeauthor{willmer-98} (circle), and \citeauthor{norb-02} (triangle).
 \label{fig:ampEfancy}
}
\end{figure}

Figure~\ref{fig:ampEfancy} shows the same data as 
Figure~\ref{fig:ampE} but plotted on a log-log scale. Here we assume
a correlation function slope of $\gamma=1.8$ and a flat lambda
cosmology as before. In this case, however, at each redshift
we subtract
the 
Gamma function and 
cosmological components of the two-point function (equations~\ref{eq:wtrnought}
and~\ref{eq:g}) 
from our measured value of \Aw.
The remaining \rnought- and $\epsilon$-dependent portion of the 
correlation function amplitude [$\gamma \log \rnought - (3 + \epsilon -
\gamma) \log (1+z)$] at each redshift is then plotted. 
The redshift baseline is too small 
and the uncertainties too large to solve meaningfully 
for $\rnought$ and $\epsilon$ simultaneously.
We therefore solved solely for $\rnought$.  
The best-fit physical clustering length $\rnought(\epsilon$) was found to be $4.02\pm0.22$ (-1.2), $5.25\pm0.28$ (0.0),
$6.55\pm0.36$ (1.0), and $8.17\pm0.45$ (2.0) as shown in 
Table~\ref{tab:bestro}. 
The lines superimposed onto Figure~\ref{fig:ampEfancy} show
these best-fit estimates
(solid for $\epsilon=-1.2$, dashed for $\epsilon=0$, dot-dashed
for $\epsilon=1$, and dotted for $\epsilon=2$)

Stronger constraints may be placed on the evolutionary parameter
$\epsilon$ if we consider our
data in conjunction with local measurements.
The \citeauthor{norb-02} measurement of $5.7\pm0.6$ ($\Mstar-0.5< M < \Mstar+0.5$)
is that which most closely resembles our own study 
with regards to the range of absolute luminosity of galaxies included in the sample
($\Mstar-0.5< M < \Mstar+0.5$ c.f.  $M < \Mstar+1.0$ in this analysis).
However, we also now compare with the clustering length measured locally
by other groups for early-type galaxies (the values in parentheses
show the range in absolute magnitude). 
\cite{shep-01} obtained $5.45\pm0.28$ ($M < \Mstar-1.0$) from the
CNOC2 survey.
However, other groups have obtained 
slightly larger correlation lengths.  \citet{love-95} measured
$6.4\pm0.7$ ($\Mstar-0.5< M < \Mstar+0.5$) for the Stromlo-APM survey,
\citet{willmer-98} measured $6.06\pm0.39$ ($M < \Mstar+5.5$) for the
SSRS2 survey
and \citet{guzzo-97} measured
$8.35\pm0.75$ for the
Perseus-Pisces survey.
Note that the latter (\citeauthor{guzzo-97}) result is for a survey
known to contain a high abundance of local clusters and therefore the
high correlation length is unsurprising. 
Also plotted  onto Figure~\ref{fig:ampEfancy} are the values 
measured by \citeauthor{love-95} (diamond), \citeauthor{guzzo-97} (star), 
\citeauthor{willmer-98} (circle) and \citeauthor{norb-02} (triangle).
Even with the above-mentioned caveats, it would appear 
from 
Figure~\ref{fig:ampEfancy} that in order to match the ``local'' value
for early-type galaxies, the correlation length of such galaxies
would be required to grow with time ($\epsilon\ga0.0$). 

Equation~(\ref{eq:xrz}) is a purely empirical model and is not necessarily
expected to be valid at all redshifts. In particular, this model
predicts monotonic evolution in \rnought\ with redshift, which is
at odds with predictions (although these provide
rather weak constraints at present) from  N-body simulations.
For example, from their semi-analytic model \citet[their Figure
2]{baugh-99} suggest that $\epsilon =0$ is a good descriptor of the 
behavior of the galaxy correlation function between $z=0$ and $z=1$
\ie\ clustering does not evolve in physical coordinates and there is a monotonic 
decrease in the physical correlation length with lookback time during
that epoch. 
However, between $z=1$ and $z=2$, a minimum in
the correlation length is reached.
At higher redshift, \citeauthor{baugh-99} predict 
an upturn in the correlation length,
and thus equation~(\ref{eq:xrz}) would not describe 
galaxy clustering well in the higher redshift regime.
Similar conclusions  were reached from simulations by \cite{kauffhiz-99}.
Such predictions of an upturn in the correlation length
at higher redshift 
are supported by the $z\sim3$ measurements of Lyman break galaxies
\citep{adel-98, giav-98} 
Thus, while the $\epsilon$ model does appear to be in qualitative
agreement with our
measurements of early type clustering at $z<1$, any evolution in
clustering at higher redshift may well be poorly described
by the $\epsilon$ formalism.
For now, based on our measurements, the predictions from semi-analytic modelling, and 
the requirement to
match the clustering lengths measured locally for early-type galaxies,
we assume the ``null hypothesis'', namely, that clustering
is fixed in physical coordinates. We conclude, for now at least, that 
the $\epsilon$ model with $\epsilon=0.0$, and a present-day
correlation length
$\rnought = 5.25\pm0.28$ provides
a satisfactory fit to observations over the redshift range
$0<z<0.9$.
(See \citet{hcb-00} for a similar conclusion of minimal evolution
relative to stable clustering in the redshift range $0<z<1$ although 
in that case comparing heterogeneous samples
of galaxies with respect to absolute magnitude, redshift, morphological type,
correlation function slope $\gamma$, and cosmology).
Note that for a correlation function of the form 
$\xr = {\left( {r}/{\rnought} \right)}^{-\gamma}$, clustering density
falls with radius as $r^{-\gamma}$ and the cosmic mean rises with
redshift as $(1+z)^{3}$, so the stable clustering prediction is for
the proper correlation length to decrease with increasing redshift as
$\rnought(z)=\rnought(0)(1+z)^{-3/\gamma}$.

Interestingly, we note that the clustering strength we measure is significantly 
weaker than that found for extremely red objects (EROs) at $z>1$.
 \citet{daddi-00,daddi-01} used $K$ and $R$ band imaging, \citet{mccarthy-01}
used $H$ and $I$ band imaging, and \citet{firth-02} used $R$ and $H$
band imaging all from the Las Campanas Infrared Survey, to
select EROs (which are thought
to be predominantly elliptical galaxies in the redshift range
$1<z<1.5$). 
\citeauthor{daddi-01}
measured a \emph{comoving} correlation length
of $12\pm3\hMpc$ for an effective redshift $z\sim1.2$. 
\citeauthor{mccarthy-01} measured a slightly smaller
\emph{comoving} correlation length of $9.5\pm0.5 \hMpc$. Using our
preferred $\epsilon=0$, 
these comoving values can be converted to physical correlation
lengths via $\xnought = \rnought (1+z)^{-(3+\epsilon-\gamma)/\gamma}$
where $\rnought = x_{0}(z=0)$ which translates to
$\rnought \sim 12-18\hMpc$. Even if one assumed that these galaxies
underwent no evolution in comoving clustering ($\epsilon=-1.2$ \ie\
$x_{0} = x_{z}$) this would result in a present-day correlation length of $\rnought \sim 9-12\hMpc$.
Even allowing for these two samples containing brighter  galaxies
than ours 
it would be very difficult to reconcile such large
correlation lengths for the ERO's with our measurements for $0<z<1$ early
types.  Indeed, it would be difficult to reconcile such high
values of correlation length with the $8.33\pm1.82$ found locally for 
the \emph{brightest} interval in the 
\citet[(their Table~2)]{norb-02} sample.
\citeauthor{firth-02} measure  
$\rnought=7.7\pm2.4$ 
 for $\epsilon=-1.2$ or $\rnought=12.1$
assuming $\epsilon=0$.
It is
possible that the width of the redshift distributions
$N(z)$ (derived from photometric redshifts) used in the Las Campanas
estimates have been overestimated leading to an increase in the
estimates of \rnought. 
Alternatively, the larger inferred values of $\rnought$ may be the result of
an increasing biasing with redshift \citep{mowhite-96}.
A yet further possibility is that ERO's are not truly field early types but may
show stronger clustering strengths because they are cluster
ellipticals
in the process of forming \citep{moussom-02}. Further data will be
required to resolve this controversy.






\section{CONCLUSIONS}
\label{sec:conc}

We investigated the dependence of the two-point galaxy angular correlation
function 
on median magnitude, $V-I$ color and morphology.
We found \wt\ to
be consistent with a power-law of slope $-0.8$
for both $I$ and $V$ passbands, down to our faintest limits of $I=24$ and $V=25$.
We found \Aw, the amplitude of \wt\ at $1\amin$,
to decrease monotonically with increasingly faint median magnitude.
We used spectroscopic redshifts measured from Cowie's SSA22 
field sample to model the galaxy redshift distribution 
as a function of apparent magnitude. We
compared the measured values of clustering amplitude \Aw\ with
the values predicted in an $\Omega_{{\rm m}0} = 0.3, 
\Omega_{\lambda 0} = 0.7$ cosmology.
We found that simple redshift-dependent models with evolutionary
parameter $\epsilon$ were inadequate to describe the evolution of 
clustering. We concluded that allowance must be made for the 
increasing proportion of later type and fainter galaxies (with weaker
correlation strengths) entering our sample at fainter magnitudes.

We also found a strong clustering dependence on color. 
Extremely blue galaxies ($V-I\sim0.5$) were found to have a clustering
amplitude 
about $15-20$ times as high as the full field population. 
This is most likely to be because many of these
blue galaxies are situated at very similar relatively low redshift, and
therefore \wt\
is minimally diluted by projection effects.
Extremely red galaxies  ($V-I\sim3$) were found to have an clustering amplitude
about $10$ times as high as the full sample. We similarly interpreted the stronger 
clustering amplitude  for redder galaxies to be mainly due to these
galaxies occupying a narrow range in redshift at $z=1$. The
stronger signal is also due in part to these being 
early-type galaxies which locally cluster 
more strongly than later types.

We then presented the first attempt to investigate redshift evolution
utilizing a population of galaxies of the \emph{same} absolute luminosity 
and morphological type.
We used $V-I$ color to isolate a sample of
early-type galaxies and investigated the  evolution in their clustering.
By making an \emph{identical} cut in absolute magnitude 
to our early-type sample at each redshift, we determined \wt\ for galaxies
with effective 
luminosity $\Leff\simeq\Lstar$ (assuming no evolution in 
the luminosity function with redshift) in 
eight redshift intervals spanning $z=0.2-0.9$. 
Although uncertainties were large,
we found the evolution in the clustering of these galaxies to be
consistent with stable clustering ($\epsilon=0$).
We found  \Lstar\ early-type galaxies to have correlation length
$\rnought = 5.25\pm0.28 \hMpc$
(assuming $\epsilon=0$), a slightly 
higher correlation length than has been found for the local full field population.
Our measured value of \rnought\
is in good agreement with the 2dFGRS measurement 
of correlation strength
for \Lstar\ early-type galaxies in the local universe. 

Over the last few years it has become increasingly apparent that
galaxy clustering has a bivariate dependence on both morphological
type and intrinsic luminosity.
Clearly, if there are differences between the clustering of various
different samples of galaxies we can immediately infer that at least one
of the galaxy samples is a biased tracer of the underlying mass
distribution. 
This paper presented the first attempt to separate the relative
contributions of luminosity and type and to
investigate the evolution in clustering of a single
galaxy population.
In the future, huge quantities of new data from
galaxy redshift and large-area imaging surveys currently in progress
will allow galaxy samples to be selected more precisely by luminosity and
type.
The 2dFGRS and SDSS are in the process of 
measuring spectroscopic redshifts for millions of galaxies 
and will determine, with incredible accuracy, 
the ``local'' correlation function (both slope and
amplitude) as a function of galaxy morphological type  and absolute
luminosity. Preliminary findings have already been reported by
\citet{norb-01, norb-02, conn-02, inf-02, zehavi-02, dodelson-02, teg-02}.
At higher redshift, next 
generation redshift surveys such as 
DEEP2 \citep{df-98,coil-01} or VIRMOS \citep{lefevre-01} will provide
tens of thousands of galaxy redshifts.

In the more immediate future, large multi-passband surveys
such as the Deep Lens Survey 
\url{(http://dls.bell-labs.com/)} are measuring tens of millions
of galaxies over tens of square degrees 
which will result in much more accurate
photometric redshift determinations than possible in this study.
The greater range of absolute
luminosity then available (limited here to $M \sim M_* \pm 1$) will allow
any luminosity dependence to the early-type galaxy correlation function to be 
determined more precisely as a function of redshift. 
Moreover, increased numbers of early-type  galaxies will greatly reduce 
uncertainties in the measurement of the amplitude and slope of the 
correlation function.
Finally, the availability of more than two passbands
will also allow photometric redshifts for late-type galaxies to be
determined and a similar 
investigation to be undertaken into their clustering evolution.

\acknowledgements We thank Len Cowie for kindly allowing the 
use of the Hawaii Survey Fields dataset. It is a pleasure to 
thank H{\aa}kon Dahle, Ian Dell'Antonio, Nick Kaiser and Gerry Luppino 
for many useful discussions.
The research described in this paper was carried out, in part, by the Jet
Propulsion Laboratory, California Institute of Technology, and was sponsored
by the National Aeronautics and Space Administration.

\clearpage
\begin{deluxetable}{ccccrrcc}
\tablewidth{0pt}
\tablecaption{Field Centers and  Seeing 
\label{tab:fields}
}
\tablehead{
\colhead{Field} &
\colhead{Pointing} &
\colhead{RA (J2000)} &
\colhead{DEC (J2000)} &	
\colhead{l } &	
\colhead{b } &	
\colhead{FWHM(I)} &	
\colhead{FWHM(V)} 	
}
\startdata
Lockman	& 1	& 10:52:43.0	& 57:28:48.0 	& $149.28$	& $53.15$	& $0''.83$	& $0''.85$	\\
	& 2	& 10:56:43.0	& 58:28:48.0 	& $147.47$	& $52.83$ 	& $0''.84$	& $0''.86$ 	\\
Groth	& 1	& 14:16:46.0	& 52:30:12.0	& $96.60$	& $60.04$	& $0''.80$	& $0''.93$	\\
	& 3	& 14:09:00.0	& 51:30:00.0 	& $97.19$	& $61.57$	& $0''.70$	& $0''.85$	\\
1650	& 1	& 16:51:49.0	& 34:55:02.0 	& $57.37$	& $38.67$	& $0''.82$	& $0''.85$	\\
	& 3	& 16:56:00.0	& 35:45:00.0 	& $58.58$	& $37.95$	& $0''.85$	& $0''.72$	\\
\enddata
\end{deluxetable}

\begin{deluxetable}{cccc}
\tablewidth{0pt}
\tablecaption{Differential Cousins I-Band Counts (N deg$^{-2}$ mag$^{-1}$)
\label{tab:countsi}
}
\tablewidth{0pt}
\tablehead{
\colhead{Magnitude}  & \colhead{$\log_{10}$(N)}  & \colhead{$\sigma_{\rm high}$} & \colhead{$\sigma_{\rm low}$}
}
\startdata
            16.0  &        1.590        &  0.037         &  0.040
\\
            16.5  &        1.617        &  0.139         &  0.205
\\
            17.0  &        2.024        &  0.208         &  0.414
\\
            17.5  &        2.250        &  0.148         &  0.227
\\
            18.0  &        2.590        &  0.106         &  0.141
\\
            18.5  &        2.817        &  0.088         &  0.111
\\
            19.0  &        3.093        &  0.114         &  0.154
\\
            19.5  &        3.317        &  0.063         &  0.074
\\
            20.0  &        3.541        &  0.044         &  0.049
\\
            20.5  &        3.737        &  0.020         &  0.021
\\
            21.0  &        3.905        &  0.025         &  0.027
\\
            21.5  &        4.078        &  0.020         &  0.021
\\
            22.0  &        4.236        &  0.032         &  0.035
\\
            22.5  &        4.395        &  0.037         &  0.041
\\
            23.0  &        4.551        &  0.026         &  0.027
\\
            23.5  &        4.696        &  0.020         &  0.021
\\
            24.0  &        4.797        &  0.024         &  0.025
\\
\enddata
\end{deluxetable}

\begin{deluxetable}{cccc}
\tablewidth{0pt}
\tablecaption{Best Fits to Counts 
\label{tab:bestfits}
}
\tablewidth{0pt}
\tablehead{
\colhead{Passband}  & \colhead{Range}  & \colhead{Slope} & \colhead{Intersection}
}
\startdata
I  & [21.0--23.5] & $0.31\pm0.01$ & $-2.68\pm0.25$
\\
V  & [22.0--24.5] & $0.43\pm0.02$ & $-5.83\pm0.45$
\\
\enddata
\end{deluxetable}

\begin{deluxetable}{clll}
\tablewidth{0pt}
\tablecaption{Comparison of Number Count Slopes
\label{tab:slopes}
}
\tablewidth{0pt}
\tablehead{
\colhead{Passband}  & \colhead{Range}  & \colhead{Slope} & \colhead{Source}
}
\startdata
I  & [21.0--23.5] & $0.31\pm0.01$ & This Work
\\
I  & [20.0--24.0] & $0.35\pm0.02$ & \citet{mcc-01}
\\
I  & [21.0--25.0] & $0.33$ & \citet{met-01}
\\
I  & [22.5--25.5] & $0.31\pm0.02$ & \citet{arnn-99}
\\
I  & [19.0--22.5] & $0.34\pm0.03$ & \citet{driv-94}
\\
\\
V  & [22.0--24.5] & $0.43\pm0.02$ & This Work
\\
V  & [22.0--24.5] & $0.43\pm0.03$ & \citet{wf-97}[averaged]
\\
V  & [22.0--24.5] & $0.404\pm0.015$ & \citet{smailhyc-95}
\\
V  & [20.5--23.0] & $0.41\pm0.01$ & \citet{driv-94}
\\
\enddata
\end{deluxetable}

\begin{deluxetable}{cccc}
\tablewidth{0pt}
\tablecaption{Differential V-Band Counts (N deg$^{-2}$ mag$^{-1}$)
\label{tab:countsv}
}
\tablewidth{0pt}
\tablehead{
\colhead{Magnitude}  & \colhead{$\log_{10}$(N)}  & \colhead{$\sigma_{\rm high}$} & \colhead{$\sigma_{\rm low}$}
}
\startdata
            17.5  &        1.667        &  0.178        &  0.306
\\
            18.0  &        1.807        &  0.138        &  0.204
\\ 
            18.5  &        2.037        &  0.240        &  0.580
\\
            19.0  &        2.309        &  0.083        &  0.103
\\
            19.5  &        2.606        &  0.108        &  0.144
\\
            20.0  &        2.843        &  0.097        &  0.125
\\
            20.5  &        3.035        &  0.095        &  0.122
\\
            21.0  &        3.255        &  0.065        &  0.077
\\
            21.5  &        3.422        &  0.057        &  0.066
\\
            22.0  &        3.644        &  0.049        &  0.055
\\
            22.5  &        3.849        &  0.037        &  0.040
\\        
            23.0  &        4.059        &  0.035        &  0.038  
\\
            23.5  &        4.298        &  0.027        &  0.029
\\
            24.0  &        4.521        &  0.031        &  0.034
\\
            24.5  &        4.701        &  0.030        &  0.032
\\
            25.0  &        4.796        &  0.040        &  0.044
\\
\enddata
\end{deluxetable}

\begin{deluxetable}{ccrc}
\tablewidth{0pt}
\tablecaption{The Dependence of $\Aw(1\amin)$ on Median Magnitude
\label{tab:wt}
}
\tablewidth{0pt}
\tablehead{
\colhead{Range}  & \colhead{Median}  & \colhead{$N_{\rm gal}$} & \colhead{$\log_{10}\Aw(1\amin)$ } 
}
\startdata
$20.0  < I \leq  21.0$ & $20.60$  & $5235$  & $-1.21\pm0.08$ 
\\
$21.0  < I \leq  22.0$ & $21.60$  & $11535$ & $-1.49\pm0.04$ 
\\
$22.0  < I \leq  23.0$ & $22.59$  & $23842$ & $-1.80\pm0.04$
\\
$23.0  < I \leq  24.0$ & $23.57$  & $47141$ & $-2.19\pm0.07$ 
\\
\
\\
$21.0  < V \leq  22.0$ & $21.62$  & $2598$ & $-1.13\pm0.06$
\\
$22.0  < V \leq  23.0$ & $22.61$  & $6883$ & $-1.30\pm0.05$
\\
$23.0  < V \leq  24.0$ & $23.63$  & $19411$ & $-1.72\pm0.05$
\\
$24.0  < V \leq  25.0$ & $24.58$  & $47057$ & $-2.26\pm0.09$
\\
\enddata
\end{deluxetable}

\begin{deluxetable}{ccc}
\tablewidth{0pt}
\tablecaption{Best-fit Redshift Scale Parameter \znought\ as a Function of Magnitude
\label{tab:znought}
}
\tablewidth{0pt}
\tablehead{
\colhead{Range}  & \colhead{Raw}  & \colhead{Corrected}
}
\startdata
$20.0  < I \leq  20.5$ & $0.15$ & $0.15$ 
\\
$20.5  < I \leq  21.0$ & $0.20$ & $0.20$ 
\\
$21.0  < I \leq  21.5$ & $0.21$ & $0.22$ 
\\
$21.5  < I \leq  22.0$ & $0.22$ & $0.26$ 
\\
$22.0  < I \leq  22.5$ & $0.26$ & $0.36$
\\
$22.5  < I \leq  23.0$ & $0.30$ & $0.41$ 
\\
$23.0  < I \leq  23.5$ & $0.34$ & $0.43$ 
\\
\\
\\
$21.5  < V \leq  22.0$ & $0.13$ & $0.13$ 
\\
$22.0  < V \leq  22.5$ & $0.15$ & $0.15$ 
\\
$22.5  < V \leq  23.0$ & $0.17$ & $0.17$ 
\\
$23.0  < V \leq  23.5$ & $0.20$ & $0.20$ 
\\
$23.4  < V \leq  24.0$ & $0.24$ & $0.26$ 
\\
$24.0  < V \leq  24.5$ & $0.30$ & $0.37$ 
\\
\enddata
\end{deluxetable}

\begin{deluxetable}{crr}
\tablewidth{0pt}
\tablecaption{The Dependence of $\Aw(1\amin)$ on $V-I$ Color for Galaxies in the Range $20.0 < I \leq 23.0$
\label{tab:wt_iv}
}
\tablewidth{0pt}
\tablehead{
\colhead{Color Range}  &   \colhead{$N_{\rm gal}$} & \colhead{$\log_{10}\Aw(1\amin)$} 
}
\startdata
$0.0  < V-I \leq  0.5$  & $136$    & $0.06\pm0.45$ 
\\
$0.5  < V-I \leq  1.0$  & $4209$   & $-1.27\pm0.08$
\\
$1.0  < V-I \leq  1.5$  & $13511$  & $-1.64\pm0.08$
\\
$1.5  < V-I \leq  2.0$  & $13587$  & $-1.50\pm0.04$
\\
$2.0  < V-I \leq  2.5$  & $5757$   & $-1.20\pm0.08$ 
\\
$2.5  < V-I \leq  3.0$  & $4211$   & $-1.03\pm0.12$
\\
$3.0  < V-I \leq  3.5$  & $1833$   & $-0.78\pm0.08$
\\
\enddata
\end{deluxetable}

\begin{deluxetable}{crr}
\tablewidth{0pt}
\tablecaption{$\Aw(1\amin)$ for Early-Type Galaxies as a Function of Redshift
\label{tab:wt_E}
}
\tablewidth{0pt}
\tablehead{
\colhead{Redshift}  &  \colhead{$N_{\rm gal}$} & \colhead{$\log_{10}\Aw(1\amin)$} 
}
\startdata
$0.2\pm0.05$ & $136$   & $0.05\pm0.36$ 
\\
$0.3\pm0.05$ & $366$   & $-0.21\pm0.07$ 
\\
$0.4\pm0.05$ & $569$   & $-0.54\pm0.11$ 
\\
$0.5\pm0.05$ & $559$   & $-0.53\pm0.15$ 
\\
$0.6\pm0.05$ & $389$   & $-0.60\pm0.26$ 
\\
$0.7\pm0.05$ & $551$   & $-0.38\pm0.10$ 
\\
$0.8\pm0.05$ & $575$   & $-0.47\pm0.16$ 
\\
$0.9\pm0.05$ & $237$   & $-0.28\pm0.12$ 
\\
\enddata
\end{deluxetable}

\begin{deluxetable}{rcc}
\tablewidth{0pt}
\tablecaption{Best-Fit Physical Correlation Length $\rnought$ and
Reduced $\chi^{2}$ for Various Values of Evolutionary Parameter $\epsilon$.
\label{tab:bestro}
}
\tablewidth{0pt}
\tablehead{
\colhead{$\epsilon$}  &  \colhead{$\rnought$} & \colhead{
  $\chi^{2}/dof$}
}
\startdata
$-1.2$ & $4.02\pm0.22$ & $1.01$ 
\\
$0.0$  & $5.25\pm0.29$ & $1.59$ 
\\
$1.0$  & $6.55\pm0.36$ & $2.70$ 
\\
$2.0$  & $8.17\pm0.45$ & $4.40$ 
\\
\enddata
\end{deluxetable}

\end{document}